\documentclass[runningheads]{llncs}

\usepackage{eccv}

\usepackage{tabularx}
\usepackage{diagbox}

\usepackage{eccvabbrv}

\usepackage{graphicx}
\usepackage{booktabs}

\usepackage[accsupp]{axessibility}  

\usepackage{hyperref}

\usepackage{orcidlink}

\definecolor{myred}{rgb}{.8,.0,.0}

\begin{document}

\title{Robustness of transferability estimation metrics for medical imaging} 


\ifdefined\DOUBLEBLIND
    \author{***}
    \authorrunning{***}
    \institute{***}
\else
    \author{Niclas Cla{\ss}en \and
    Théo Sourget \and
    Dovile Juodelyte \and
    Rob van der Goot \and
    Veronika Cheplygina \\
    }
    \authorrunning{N. Claßen et al.}
    \institute{
    IT University of Copenhagen, Denmark\\
    \email{\{niclc,vech\}@itu.dk}\\
    }
\fi

\authorrunning{N.~Cla{\ss}en et al.}


\maketitle

\begin{abstract} 
In transfer learning, the choice of source model largely influences the performance on a target dataset. Still, selecting a fitting source remains a challenging task, especially in medical imaging where one has to decide between models pre-trained on off-the-shelf options, such as ImageNet, and domain specific datasets. Transferability estimation (TE) metrics address this problem by aiming to predict the best performing source model in a computationally cost effective way. However, previous work has reported
conflicting TE metric performances due to differences in experimental setups. Moreover, most TE metrics are designed for and evaluated on natural images, while being optimized for accuracy, whereas in medical imaging metrics that are more robust to class imbalance are typically used. We study the impact of varying the target dataset as an isolated factor, by constructing miniature populations of different sample sizes and random seeds. In addition, we investigate the influence of the evaluation metric used to obtain the reference ranking. We find that small modifications to the target dataset change the rankings. Furthermore, we show that the choice of evaluation metric affects the reference rankings and therefore the evaluation of TE metrics. Overall, we observe a low agreement between rankings from TE metrics and reference. The code, model checkpoints and data splits used in this work are available through \href{https://github.com/niclasclassen/robustness-of-transferability-estimation-metrics-for-medical-imaging}{GitHub}.



\keywords{Transfer Learning \and Transferability estimation \and Medical imaging \and Robustness}
\end{abstract}

\section{Introduction}
\label{sec:intro}
Transfer learning (TL) is one of the most commonly used training paradigms for medical image classification, because of its inherent data scarcity problems. TL offers a solution to this by leveraging information from a source domain with more easily obtainable data to improve performance on a target task where data is scarce \cite{cheplyginaNotsosupervisedSurveySemisupervised2019,kimTransferLearningMedical2022,ataseverComprehensiveSurveyDeep2023}. Nevertheless, it remains unclear what source is best suited for a given target task. In previous work, natural image datasets and especially ImageNet \cite{dengImageNetLargeScaleHierarchical} are commonly preferred over in-domain datasets \cite{ataseverComprehensiveSurveyDeep2023}. At the same time, other research suggests that natural images may not always be the optimal source for medical image classification tasks, as pathology is often indicated by subtle local texture variation \cite{raghuTransfusionUnderstandingTransfer2019}. Furthermore, some prior work indicates that TL is most effective when source and target tasks are in the same image domain \cite{mensinkFactorsInfluenceTransfer2021,kronesPretrainingDownstreamPerformance2026}. Often, several backbone models are empirically evaluated \cite{kimTransferLearningMedical2022} in experimental setups. However, comparing multiple source configurations with each other is not only computationally expensive and time consuming, it may fail to identify superior source candidates, as only a limited set of options can be considered.

Transferability estimation (TE) metrics \cite{nguyen_leep_nodate,bao_information-theoretic_2022,bolya_scalable_nodate,juodelyte_dataset_2024,li_ranking_nodate,you_logme_nodate,avidan_not_2022,wang_how_2023,tran_transferability_2019,pandy_transferability_2022} address this problem by estimating how well a source will perform on a target task without the need to extensively fine-tune all of them. This leads to a shift from choosing the optimal source configuration to selecting the right TE metric. However, previous work \cite{agostinelli_how_2022} indicates that small variations to an experimental setup result in different conclusions about the superiority of a TE metric over another. Their experiments show that the choice of target dataset has the biggest impact followed by the evaluation measure. Still, there are no clear requirements defined that a target dataset needs to fulfill for the TE metrics to be reliably applicable. Moreover, most TE metrics are designed for and evaluated on natural images, while being optimized for accuracy. Not surprisingly, Chaves et al. \cite{chaves_performance_2023} show that current TE metrics do not reliably transfer in a medical context, where Area under the receiver operating characteristic curve (AUROC) is typically used over accuracy as a more robust metric. 

\begin{figure}[h]
    \centering
    \includegraphics[width=0.85\textwidth]{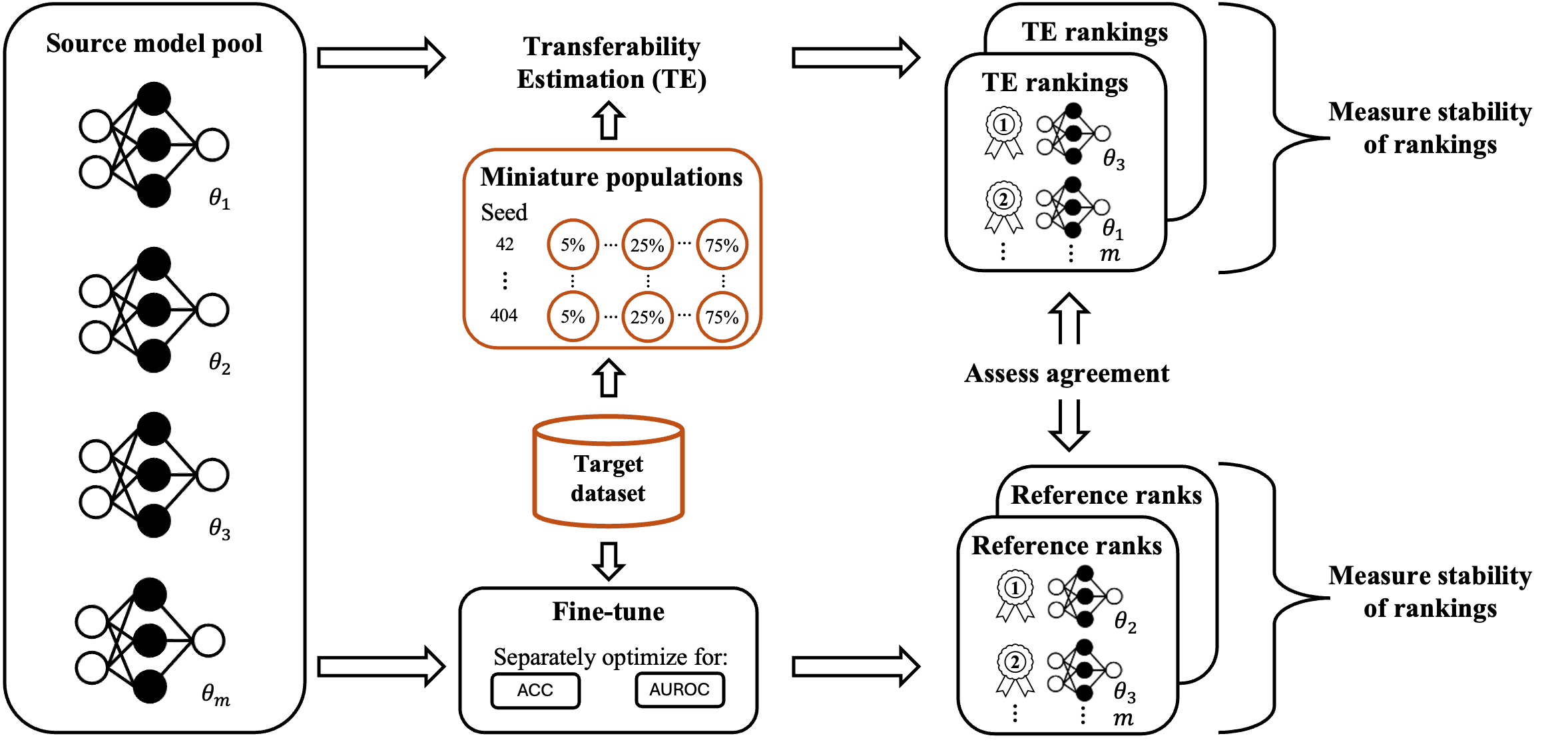}
    \caption{Illustration of our study. We asses the robustness of existing TE metrics with regards to different target dataset representations, as well as the evaluation metric used for the reference ranking.}
    \label{fig:experimental_setup}
\end{figure}
 
In this work, we take a step back and focus on the robustness of existing TE metrics. Figure \ref{fig:experimental_setup} presents an overview of our study design.  Our contributions are threefold: (1) We find that TE metrics are sensitive to small variations of the same target dataset. (2) We show that the choice of evaluation metric affects the reference ranking. (3) We provide benchmark results for MedMNIST \cite{yang_medmnist_2023} that allow future comparisons.



\section{Related Work}
\label{sec:related_work}


\subsection{Transferability estimation metrics}

Following Juodelyte et al. \cite{juodelyte_dataset_2024}, we group existing TE metrics based on whether they evaluate the initial fit of source models from static features or aim to model dynamics during fine-tuning. 

\subsubsection{Static feature evaluation:}
Tran et al. \cite{tran_transferability_2019} introduced a model-agnostic approach using an information-theoretic measure, which quantifies the information required to estimate a label in one task given the known label in another. Nevertheless, this method is limited to tasks where source and target share the same input instances. Building on this, Nguyen et al. \cite{nguyen_leep_nodate} presented the \textit{Log Expected Empirical Prediction} (\textit{LEEP}), which constructs an empirical predictor from the joint distribution of source model predictions and target labels, computing the log-likelihood of target labels given source predictions. They demonstrate that \textit{LEEP} remains effective in small and imbalanced data settings, evaluating correlation against test accuracy (or F1 score in the imbalanced case). However, their imbalanced data experiments are restricted to binary classification with a fixed 1:5 class ratio, leaving the metric's behavior under more varied imbalance and class count settings untested. Furthermore, the impact of the evaluation metric used for the reference ranking remains unexplored. Li et al. \cite{li_ranking_nodate} extended \textit{LEEP} and introduced \textit{Gaussian LEEP} ($\mathcal{N}\textit{LEEP}$) to support pre-trained models that lack a classification head, by fitting a Gaussian mixture model to the target embeddings in place of the source classification head. 


Pándy et al. \cite{pandy_transferability_2022} proposed another approach called the \textit{Gaussian Bhattacharyya Coefficient} (\textit{GBC}), which models target classes as per-class Gaussians in the source embedding space and measures their pairwise overlap: lower overlap indicates better expected transferability. Bolya et al. \cite{bolya_scalable_nodate} introduced \textit{Pairwise Annotation Representation Comparison (PARC)}, which scores source models based on the Spearman correlation between two distance matrices for all pairs of images, where one matrix is based on target images in the feature space of the source model and the other between the target labels. 
Bao et al. \cite{bao_information-theoretic_2022} proposed a different method, named \textit{H-score}, which is based on the intuition that a model transfers well to a target dataset if the target embeddings have high inter-class variance and low feature redundancy. These quantities are computed by constructing the inter-class and data covariance matrices. 


\subsubsection{Modeling dynamics during fine-tuning:}
Instead of evaluating static features, a slightly different branch of research attempts to approximate the dynamics of fine-tuning. You et al. \cite{you_logme_nodate} introduced \textit{LogME}, which estimates the maximum log-evidence of target labels given source features via a Bayesian linear model. 
Shao et al. \cite{avidan_not_2022} proposed a method called, \textit{Self-challenging Fisher Discriminant Analysis} (\textit{SFDA}), which projects source features into a Fisher space to enhance class separability while a self-challenging mechanism focuses discrimination on hard examples. Wang et al. \cite{wang_how_2023} presented the \textit{Neural Collapse Transferability Index} (\textit{NCTI}), motivated by the neural collapse phenomenon at the terminal stage of training. \textit{NCTI} measures how far the source model's target features are from the neural collapse state, combining within-class variability collapse, simplex encoded label interpolation geometry, and nearest-centroid classifier applicability. Additionally, they investigate how transferability estimation degrades under limited target data, by randomly sampling between 2 and 500 images per class and measuring the resulting ranking correlation. They observe a general performance degradation for smaller sample sizes across methods. However, this experiment is run on only one dataset with a single random sample per size, leaving open how much the result would vary under repeated resampling.

\subsection{Transferability estimation in medical imaging}

Despite the prevalence of transfer learning in medical image analysis, transferability estimation has received limited attention in this domain, and the methods discussed above were not evaluated on medical targets, with the exception of $\mathcal{N}\textit{LEEP}$, which included PatchCamelyon \cite{veeling2018rotation} as one of four downstream tasks. To address this gap, Chaves et al. \cite{chaves_performance_2023} conducted a systematic evaluation of seven TE metrics across three medical classification tasks: melanoma detection, breast cancer histopathology, and brain tumor classification. They found that no existing metric can reliably and consistently predict target performance in medical imaging contexts. Their results show that metrics performing well on general-purpose datasets fail under the domain shift from natural to medical images. However, the authors mention themselves that further work is needed to evaluate the robustness of TE metrics under limited data availability, and class imbalance. Furthermore, they use balanced accuracy in their experiments, while most TE metrics are designed to match the reference performance ranked by accuracy. This leaves open whether their conclusions result from the medical data or the evaluation metric used for the reference ranking.

Juodelyte et al. \cite{juodelyte_dataset_2024} similarly demonstrated that TE metrics designed and validated on natural image datasets perform poorly in medical image classification, and established a transfer performance benchmark across 15 source datasets and 9 CNN architectures on 11 MedMNIST \cite{yang_medmnist_2023} target tasks. Furthermore, they propose \textit{LPFU}, a transferability metric that combines feature quality with gradients to evaluate both the initial suitability and potential adaptability of source models to a given target task. However, similar to Chaves et al. \cite{chaves_performance_2023}, they use a different evaluation metric for the reference ranking, without investigating its impact. Moreover, the experiments are performed on a fixed subsets size per target, created with a single random seed, which limits the reliability of these results.




\subsection{Issues with existing transferability estimation methods}

Despite the growing number of proposed metrics, several systematic weaknesses have been documented. A fundamental concern is evaluation instability across experimental setups. Agostinelli et al. \cite{agostinelli_how_2022} conducted a large-scale study comprising 715k experimental variations spanning different source model pools, target datasets, and evaluation measures. They found that even small changes in any of these components lead to different conclusions about which metric is superior, with the target dataset and evaluation metric having the biggest impact. 
Adding to this, previous work commonly simulates TL settings by creating subsets of larger target datasets. However, this is often performed with a single random seed, raising questions about the robustness of these results.

Statistical estimation failures under limited data are a separate but related concern. Ibrahim et al. \cite{ibrahim_newer_2023} showed that \textit{H-score} is unreliable when the number of target samples is small, because estimating its two covariance matrices becomes ill-conditioned when the feature dimension greatly exceeds the sample count. They proposed a shrinkage-based regularization that improves rank correlation. 
Similarly, \textit{SFDA} is prone to overfitting in low-sample regimes: when the feature dimension exceeds the number of target samples per class, the within-class scatter matrix becomes rank-deficient, and the learned projection can collapse all samples of a class onto a single point rather than approximating fine-tuning dynamics \cite{juodelyte_dataset_2024}. 


Collectively, these findings point to a shared vulnerability: TE metric rankings are sensitive to the quantity and composition of the target data used for metric computation. This sensitivity has not been systematically studied as an isolated variable. Furthermore, the impact of the evaluation metric used for the reference ranking remains unexplored. The present work addresses this gap by (1) benchmarking how rankings produced by a broad set of transferability metrics change as a function of target dataset subset size and sampling, and (2) evaluating the impact of the evaluation metric used for the reference ranking using medical imaging target datasets where data scarcity is a practical constraint.

\section{Methods}
\label{sec:methods}

\subsection{Problem definition}

We consider a set of $m$ source models $\mathcal{S} = \{ \theta_1, \theta_2, \ldots, \theta_m \}$ and a target dataset $T = \{(\mathbf{x_i}, y_i)\}_{i=1}^{n}$ with $n$ labeled data points. TE metrics aim to rank the source models within $S$ according to their performance on $T$ after fine-tuning. This is done by applying a scoring function to estimate the transferability. Importantly, the computational cost of these TE metrics is significantly lower than that of fine-tuning every model. In this work, we focus on how varying target representations $t$, where $t \subseteq T$, as well as the choice of evaluation metric used for the reference ranking affect current TE metrics. Notably, we compare the full rankings of source models instead of only considering the top-ranked ones, providing a more complete assessment. Identical scores are assigned the same ranks, with subsequent ranks determined according to standard competition ranking (e.g., 1, 2, 2, 4). 


\subsection{Data}
To evaluate the robustness of existing TE metrics, we use MedMNIST v2 \cite{yang_medmnist_2023}, a large-scale dataset collection of standardized twelve 2D and six 3D biomedical image datasets for classification tasks. Given that most TE metrics are designed for 2D binary or multi-class classification, we focus purely on the 2D 224x224 datasets of which we exclude four (two because they are not binary or multi-class classification tasks and two due to limited computational resources). This leaves us with eight potential targets. However, all 2D datasets, except the target itself, are part of the source pool. Beyond MedMNIST v2 we include ImageNet \cite{dengImageNetLargeScaleHierarchical} a large-scale natural image dataset widely used for pre-training. An overview of these datasets is shown in Table \ref{tab:overview_data}.

\subsection{Experiments}


\subsubsection{Impact of target dataset representation (Ex1):} In our first experiment, we explore how rankings from TE metrics behave across a variation of representative target subsets. It should be noted that the term representative subset can be understood in different ways as described by Clemmensen et al. \cite{clemmensen_data_2023}. In our work, a subset is considered representative if it mimics the class label distribution of the reference population. This follows the notion of a miniature population for which the representativeness increases with the subset size. 


\begin{table}[!ht]
\centering
\caption{Target and source datasets used in our experiments, consisting of MedMNIST v2 \cite{yang_medmnist_2023} and ImageNet \cite{dengImageNetLargeScaleHierarchical}. The datasets are sorted by the number of train images in ascending order.}
\label{tab:overview_data}
\setlength{\tabcolsep}{3pt}

\begin{tabularx}{\linewidth}{l|c r r r r}
\toprule
\textbf{Dataset} &
\textbf{Modality} &
\multicolumn{1}{c}{\textbf{\#Classes}} &
\multicolumn{3}{c}{\textbf{\#Images}} \\
\cmidrule(lr){4-6}
&
&
&
\multicolumn{1}{c}{\textbf{Train}} &
\multicolumn{1}{c}{\textbf{Validation}} &
\multicolumn{1}{c}{\textbf{Test}} \\

\midrule
\multicolumn{6}{c}{\textbf{Source and Target Datasets}}\\
\midrule

Breast & Breast Ultrasound & 2 & 546 & 78 & 156 \\
Pneumonia & Chest X-Ray & 2 & 4,708 & 524 & 624 \\
Derma & Dermatoscope & 7 & 7,007 & 1,003 & 2,005 \\
Blood & Blood Cell Microscope & 8 & 11,959 & 1,712 & 3,421 \\
OrganC & Abdominal CT & 11 & 12,975 & 2,392 & 8,216 \\
OrganS & Abdominal CT & 11 & 13,932 & 2,452 & 8,827 \\
OrganA & Abdominal CT & 11 & 34,561 & 6,491 & 17,778 \\
OCT & Retinal OCT & 4 & 97,477 & 10,832 & 1,000 \\

\midrule
\multicolumn{6}{c}{\textbf{Source-only Datasets}}\\
\midrule
Retina & Fundus Camera & 5 & 1,080 & 120 & 400 \\
Chest & Chest X-Ray & 14 & 78,468 & 11,219 & 22,433 \\
Path & Colon Pathology & 9 & 89,996 & 10,004 & 7,180 \\
Tissue & Kidney Cortex Microscope & 8 & 165,466 & 23,640 & 47,280 \\
ImageNet & Natural Images & 1,000 & 1,281,167 & 50,000 & 100,000 \\

\bottomrule
\end{tabularx}
\end{table}

\textbf{Miniature populations:} We consider the target datasets shown in Table \ref{tab:overview_data} as our reference populations. Based on these, we construct miniature populations using 5\%, 10\%, 25\%, 50\%, and 75\% fractions of the targets train data. The sampling is stratified on the class label to ensure equal class distributions and uses nested subsets, meaning that each smaller subset is contained within the larger ones. We repeat this sampling procedure five times using different random seeds.

\begin{figure}[h]
    \centering
    \includegraphics[width=0.83\textwidth]{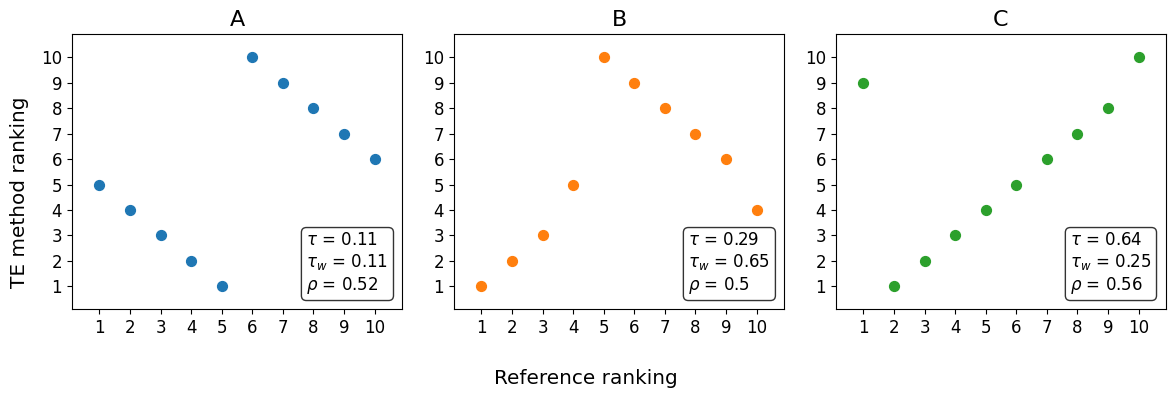}
    \caption{Comparing Spearman's rank correlation coefficient $\rho$, Kendall's Tau $\tau$, and weighted Kendall's Tau $\tau_w$ under different ranking scenarios. In A we have a moderate agreement according to $\rho$, while $\tau$ and $\tau_w$ indicate a weak correlation. Moreover, C is superior over B when looking at $\tau$, whereas B outperforms C when considering $\tau_w$. All correlation coefficients range from $-1$ to $1$, where $-1 =$ perfect negative correlation, $0 =$ no correlation, and $1 =$ perfect positive correlation.}
    \label{fig:correlation_coeff_diffs}
\end{figure}

\textbf{Measure of correlation:} In previous work, different correlation coefficients are used to evaluate the performance of TE metrics, by assessing their agreement with a reference ranking. Typically, this includes Spearman's rank correlation coefficient $\rho$ \cite{spearmanProofMeasurementAssociation1904}, Kendall's Tau \cite{kendallNewMeasureRank} $\tau$, and weighted Kendall's Tau \cite{loshchilovDecoupledWeightDecay2019} $\tau_w$. More recently, weighted Kendall's Tau is often preferred over others, as it assigns higher importance to the top-ranked source models. This seems logical, but favors experimental setups where one source model is dominating while the other ranks could all be mixed up. However, the existence of a superior source model is not guaranteed. In general, the choice of correlation coefficient can lead to different interpretations of the results, affecting both the assessment of alignment with the reference ranking and the relative superiority of one TE metric over the others, as shown in Figure \ref{fig:correlation_coeff_diffs}. For a more robust analysis, we consider all three correlation coefficients in our work.

\textbf{Measure of stability:}
We assess the stability of existing TE metrics from two complementary perspectives. First, we measure the stability of TE metrics across subsets of equal size generated with different random seeds. We denote by $r$ the rankings of source models from a single TE metric. Let $t_i$ and $t_j$ refer to target dataset representations of the same fraction but generated with different random seeds, $u$ denotes the number of unique pairs for which $i$ and $j$ are the indices. The correlation coefficient is denoted as $corr$. We then define the intra-metric stability as follows:

\begin{equation}
\label{eq:stability_intra}
stability_{intra}(r, T)
=
\frac{1}{u}
\sum_{i=1}^{u-1}\sum_{j =i+1}^u
corr(r(t_i), r(t_j))
\end{equation}

Second, we explore the generalizability of TE metrics by assessing the stability between rankings from subsets with the ranking from using the whole train data set. We denote by $t_i$ the subset generated under the $i$-th random seed and by $t'$ the subset used for the reference ranking. Let $r$ and $r'$ be two rankings of the source models, which in this case are both from a the same TE metric. Define $k$ as the number of random seeds the data was generated with. The stability of rankings w.r.t to a reference ranking is defined as follows:

\begin{equation}
\label{eq:stability_ref}
stability_{ref}(r, r', T)
=
\frac{1}{k}
\sum_{i=1}^k
corr(r(t_i), r'(t'))
\end{equation}

Additionally, we investigate the stability across TE metrics by comparing rankings between them for the same target data representations. It should be noted that a high or low level of agreement does not indicate how well a TE metric performs, but rather is used to indicate similarities or discrepancies. Let $r$ and $r'$ be two rankings from different TE metrics and $t_i$ the subset generated under the $i$-th random seed. Then, the inter-metric stability is given by: 

\begin{equation}
\label{eq:stability_inter}
stability_{inter}(r, r', T)
=
\frac{1}{k}
\sum_{i=1}^k
corr(r(t_i), r'(t_i))
\end{equation}

\textbf{Technical setup:} We use existing implementations of TE metrics from either the original papers or other related work that covers these metrics. In particular, we consider the following TE metrics: \textit{H-score, LEEP, $\mathcal{N}\textit{LEEP}$, LogME, NCTI, PARC, and SFDA}. We add minor adjustments to fit our data and ensure reproducibility. The correlation coefficients are computed via the Python library SciPy \cite{2020SciPy-NMeth}. 

\subsubsection{Influence of evaluation metric used for reference ranking (Ex2):}
To address the limitations of prior work that base the ``actual'' transferability performance solely on the accuracy of fine-tuned source models, we adopt a more robust approach. Following recommendations from Metrics Reloaded \cite{maier-heinMetricsReloadedRecommendations2024}, considering the target tasks, we evaluate each fine-tuned model not only on accuracy but also on AUROC, which is better suited for datasets with class imbalances.

\textbf{Fine-tuning:} We fine-tune all parameters of the pre-trained source models on the full train sets of our target datasets which are considered as the reference populations. We perform hyperparameter tuning for each source-target combination optimizing once for accuracy and the other time for AUROC. The final performance is evaluated on the test set resulting in our reference ranking. In addition, to see if the results from fine-tuning remain consistent across target representations, we repeat this procedure using the 5\% fractions of the target train data. This resembles a more typical transfer learning setting when data is scarce. For each target, we consider two subsets of size 5\% with different random seeds which is mainly limited by the computational resources available.

\textbf{Measure of stability:} We use Equation \ref{eq:stability_ref} to assess the agreement between TE metrics and a reference ranking, where $r$ refers to rankings from a TE metric, $r'$ to the reference ranking (once for accuracy and once for AUROC) from fine-tuning. We denote by $t'$ the data used for the reference ranking, and by $t_i$ a subset with $i$ indicating the random seed. Furthermore, we evaluate the stability of reference rankings from fine-tuning between evaluation metrics, subset sizes, and random seeds. Similarly to our previous experiment, we consider Spearman rho, Kendall's Tau and weighted Kendall's Tau as correlation coefficients.

\textbf{Technical setup}: We use ResNet-18 \cite{He2016Resnet} as model architecture, as this allows us to use the trained models from MedMNIST v2 \cite{yang_medmnist_2023} as our source candidates without the need of pre-training them. For ImageNet, we use the weights \texttt{IMAGENET1K\_V1} available via torchvision \cite{torchvision2016}. The fine-tuning is implemented using PyTorch \cite{Paszke2019PyTorch}. For hyperparameter tuning we use Optuna \cite{akibaOptunaNextgenerationHyperparameter2019} for 50 trials with a learning rate and weight decay sampled on a logarithmic scale over the ranges from $1e^{-5}$ to $1e^{-2}$ and $1e^{-6}$ to $1e^{-2}$, respectively. The batch size is set to 128. We utilize the default sampler \texttt{TPESampler} and pruning metric \texttt{MedianPruner}. As optimizer we use AdamW \cite{loshchilovDecoupledWeightDecay2019} and CosineAnnealingLR \cite{loshchilovSGDRStochasticGradient2017}. We run the experiments distributed across the following GPUs: three NVIDIA GB10, two NVIDIA A100 40GB, one A100 80GB, and one NVIDIA A30 24GB. The energy consumption while fine-tuning is tracked using Carbontracker \cite{anthonyCarbontrackerTrackingPredicting2020} and reported in Appendix \ref{appx:finetune_performance}.

\section{Results}
\label{sec:results}

\subsection{Impact of target dataset representation (Ex1)}
\label{results_intra_stability}

\subsubsection{TE metrics are sensitive to target dataset variations:} 
\begin{figure}[h]
    \centering
    \includegraphics[width=1\textwidth]{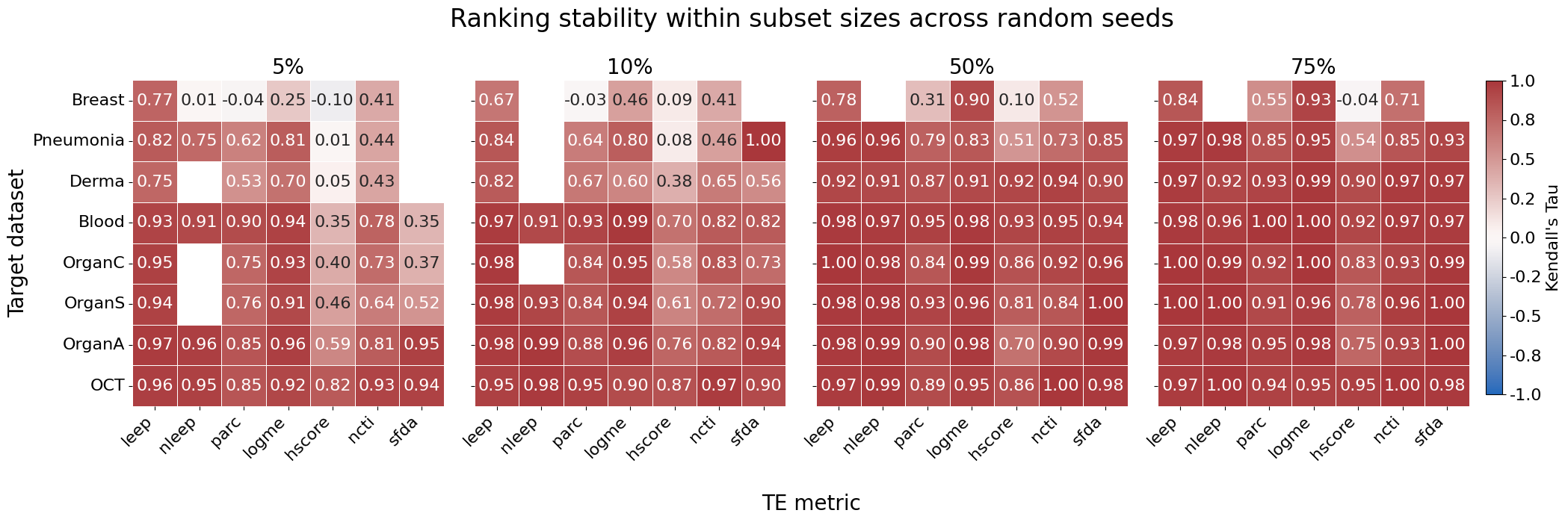} 
    \caption{Average (over 5 random seeds) pairwise Kendall's Tau ($stability_{intra}$) across target subsets of varying sizes for different TE metrics. The target datasets are sorted by absolute train set size in ascending order. Blank fields indicate NaN values.}
    \label{fig:stability_intra_kendalls}
\end{figure}

We first investigate how consistently TE metrics rank source models across different random seeds for the same subset sizes. The results with Kendall's Tau as correlation coefficient are shown in Figure \ref{fig:stability_intra_kendalls}. The outcomes for all three correlation coefficients, including Spearman's rho and weighted Kendall's Tau, are presented in Appendix \ref{appx:intra_metric_stability}. It should be noted, that $\mathcal{N}LEEP$ and \textit{SFDA} include blank fields which correspond to NaN values. For $\mathcal{N}LEEP$ this results from components with ill-defined empirical covariance when fitting the Gaussian mixture model. This effect can be mitigated by increasing the Principal component analysis energy and reducing the number of Gaussian components, as shown in Appendix \ref{appx:intra_metric_stability_modified_nleep}. For \textit{SFDA}, this is due to all source models being assigned the same score which aligns with the findings by Juodelyte et al. \cite{juodelyte_dataset_2024}, leading to NaN values when computing the correlation coefficient. As a result, the applicability of both TE metrics is limited by the target dataset.

Overall, consistent trends are observed across all three correlation coefficients. Still, for individual cases differences between these can be as large as $0.227$, highlighting the importance of considering all three for a more complete picture. Looking at how the TE metrics behave across target datasets and subset sizes, we find greater variability among smaller target representations among all TE metrics. This is especially evident for \textit{Breast}, which is about one-eight the size of \textit{Pneumonia}, the next larger target dataset. While \textit{LEEP} remains the most consistent on smaller datasets, rankings from \textit{H-score} appear to be the least stable across all target representations. A similar pattern can be observed across subset sizes compared with the rankings from using the whole train data, as shown in Figure \ref{fig:stability_ref_from_metric_k} for Kendall's Tau and for all correlation coefficients in Appendix \ref{appx:stability_ref_from_metric_all}. This suggests that the generalizability of rankings from a small subset to the broader population is limited, as the ranking changes.

\begin{figure}[h]
    \centering
    \includegraphics[width=1\textwidth]{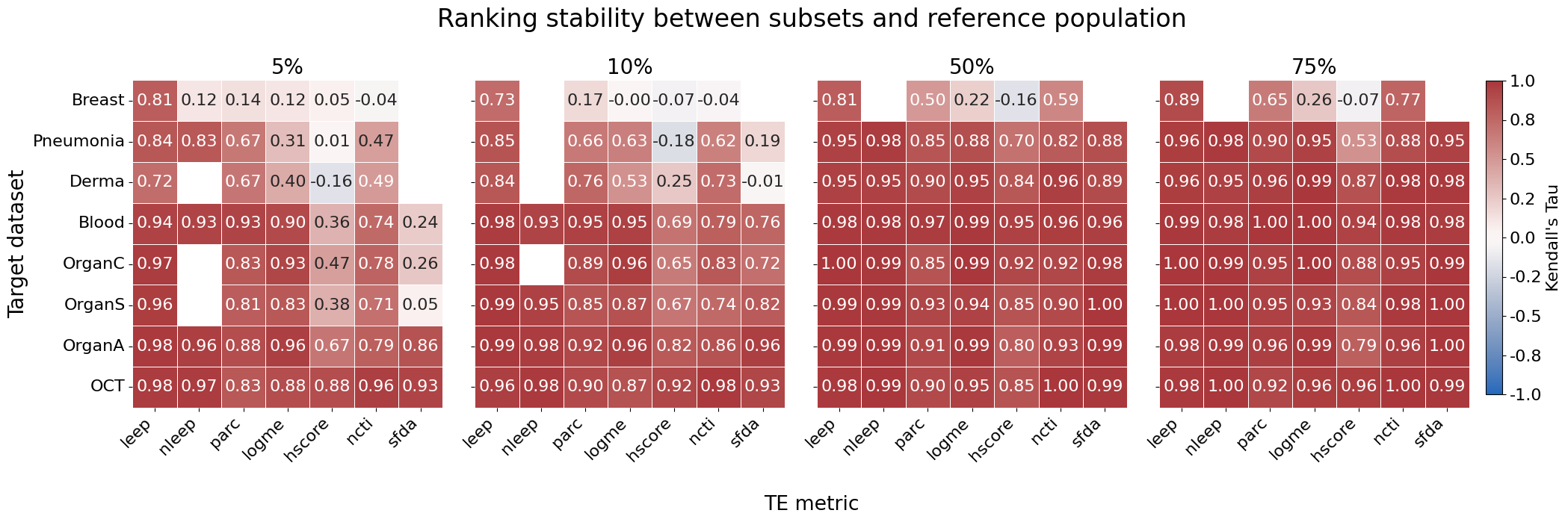} 
    \caption{Average (over 5 random seeds) pairwise Kendall's Tau ($stability_{ref}$) between rankings obtained from target subsets of varying fraction sizes and the ranking obtained using 100\% of the data. The target datasets are sorted by absolute train set size in ascending order. Blank fields indicate NaN values.}
    \label{fig:stability_ref_from_metric_k}
\end{figure}

\begin{figure}[t]
    \centering
    \includegraphics[width=1\textwidth]{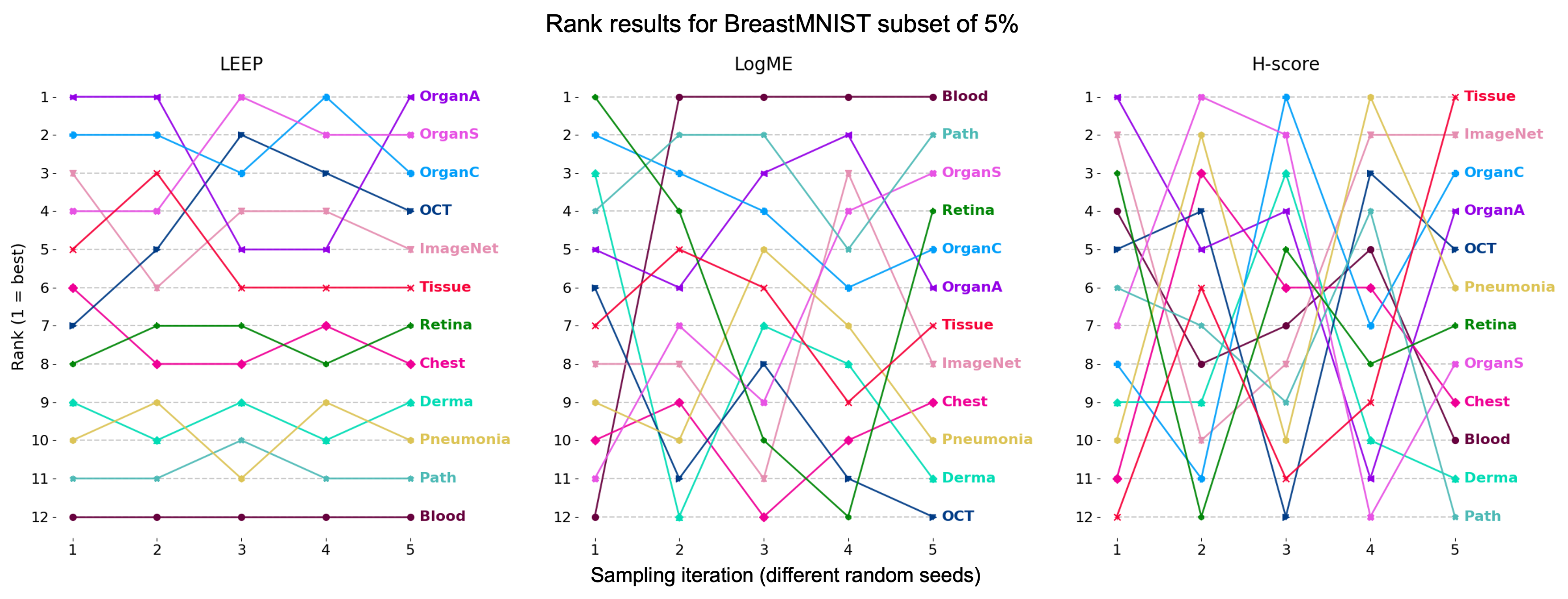} 
    \caption{Source model rankings based on TE metrics \textit{LEEP}, \textit{LogME}, and \textit{H-score} for \textit{Breast} subsets with fraction size 5\%. The subsets are generated with five different random seeds.}
    \label{fig:ranking_bump_plot}
\end{figure}

Although, \textit{LEEP} is most stable on smaller datasets, the resulting rankings still vary across random seeds, as shown in Figure \ref{fig:ranking_bump_plot}. For instance, \textit{OrganA} can be ranked first or fifth depending on the random seed for \textit{Breast} as target. This effect is more extreme for \textit{LogME} for which \textit{Blood} is ranked as both the most and least suitable source for different subsets. Moreover, we observe that rankings across TE metrics do not change consistently across random seeds. For example, \textit{LEEP} ranks \textit{Blood} last for all sampling repetitions while \textit{LogME} ranks it once last and the other times first. These findings indicate that experiments comparing TE metrics with each other are not reliable if they are based on subsets generated with a single random seed.

\subsubsection{Target specific agreement between TE metrics:}
\begin{figure}[h]
    \centering
    \includegraphics[width=0.95\textwidth]{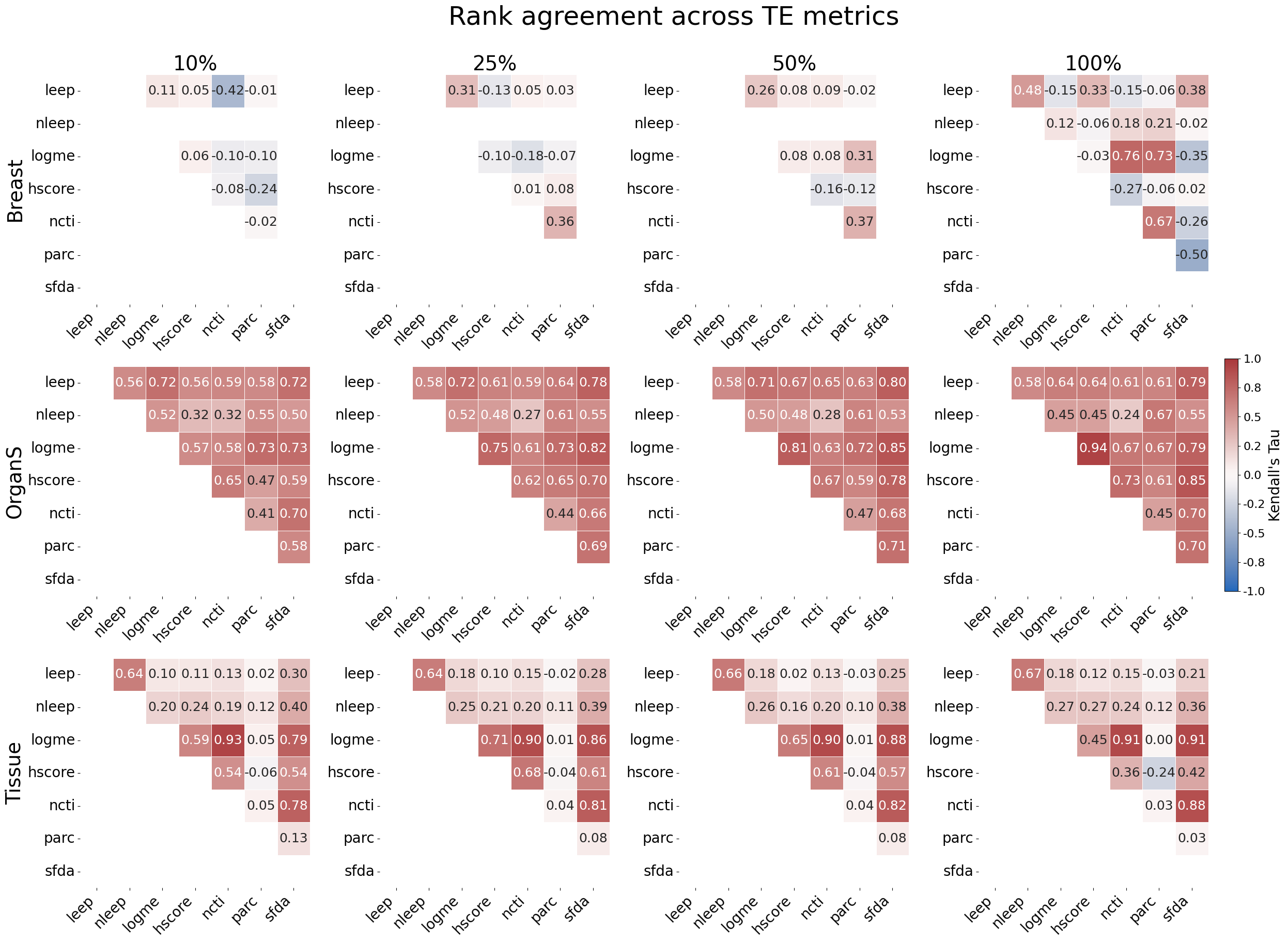} 
    \caption{Average pairwise Kendall's Tau ($stability_{inter}$) between rankings from different TE metrics. For subset sizes $< 100\%$, agreement is computed between subsets generated with identical random seeds and then averaged over 5 seeds.}
    \label{fig:stability_inter_kendalls}
\end{figure}

When we look at the agreement between TE metrics, shown in Figure \ref{fig:stability_inter_kendalls}, we notice that the rankings of source models differ especially for smaller target datasets such as \textit{Breast}. For larger datasets such as \textit{OrganS}, TE metrics rank the source models similarly. However, for the largest dataset \textit{Tissue}, the agreement between several TE metrics decreases again. This suggests that alignment between TE metrics are target dataset specific. The results for all target datasets are presented in Appendix \ref{appx:stability_inter}.

\subsection{Influence of evaluation metric used for reference ranking (Ex2)}
\label{results_reference_ranking}
\subsubsection{Reference ranking depends on evaluation metric:}
\begin{figure}[h!]
    \centering
    \includegraphics[width=0.98\textwidth]{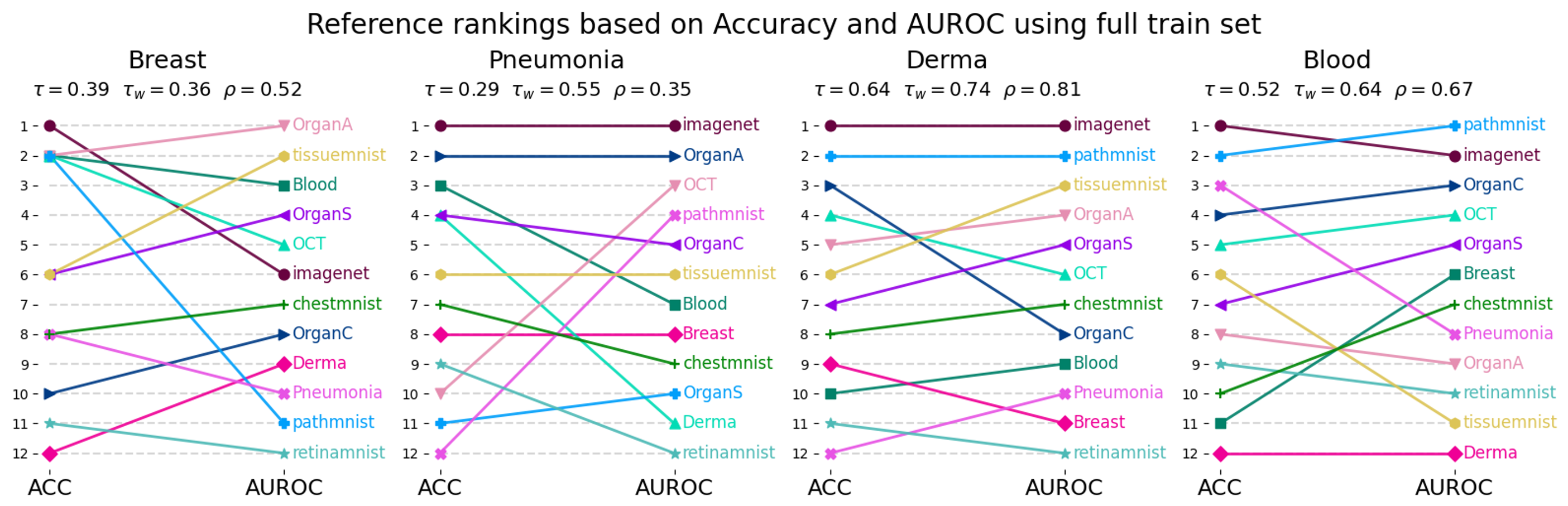}
    \caption{Stability of rankings obtained from two fine-tuning runs, optimized separately for accuracy (ACC) and AUROC on the full train set, considering Kendall's Tau $\tau$, weighted Kendall's Tau $\tau_w$, and Spearman's rho $\rho$.}
    \label{fig:source_rankings_AUC-ACC_100cpt}
\end{figure}

To investigate whether rankings change based on the evaluation metric, we fine-tune on the full target train sets while separately optimizing for accuracy and AUROC. We show resulting rankings in Figure \ref{fig:source_rankings_AUC-ACC_100cpt} for a selection of target datasets, while the results for all are included in Appendix \ref{appx:reference_ranking_alignment_auc_acc_all}. Across all target datasets, we observe that model rankings differ depending on the evaluation metric. Almost none of the source models are ranked the same across accuracy and AUROC. The choice of evaluation metric can even decide if a source model is ranked second or eleventh, as it is the case with \textit{Path} for \textit{Breast}. Therefore, it cannot be expected that current TE metrics perform well across multiple evaluation metrics, as this aspect is not incorporated into their design.


\subsubsection{Low agreement between TE metrics and reference rankings:}
\begin{figure}[h]
    \centering
    \includegraphics[width=0.98\textwidth]{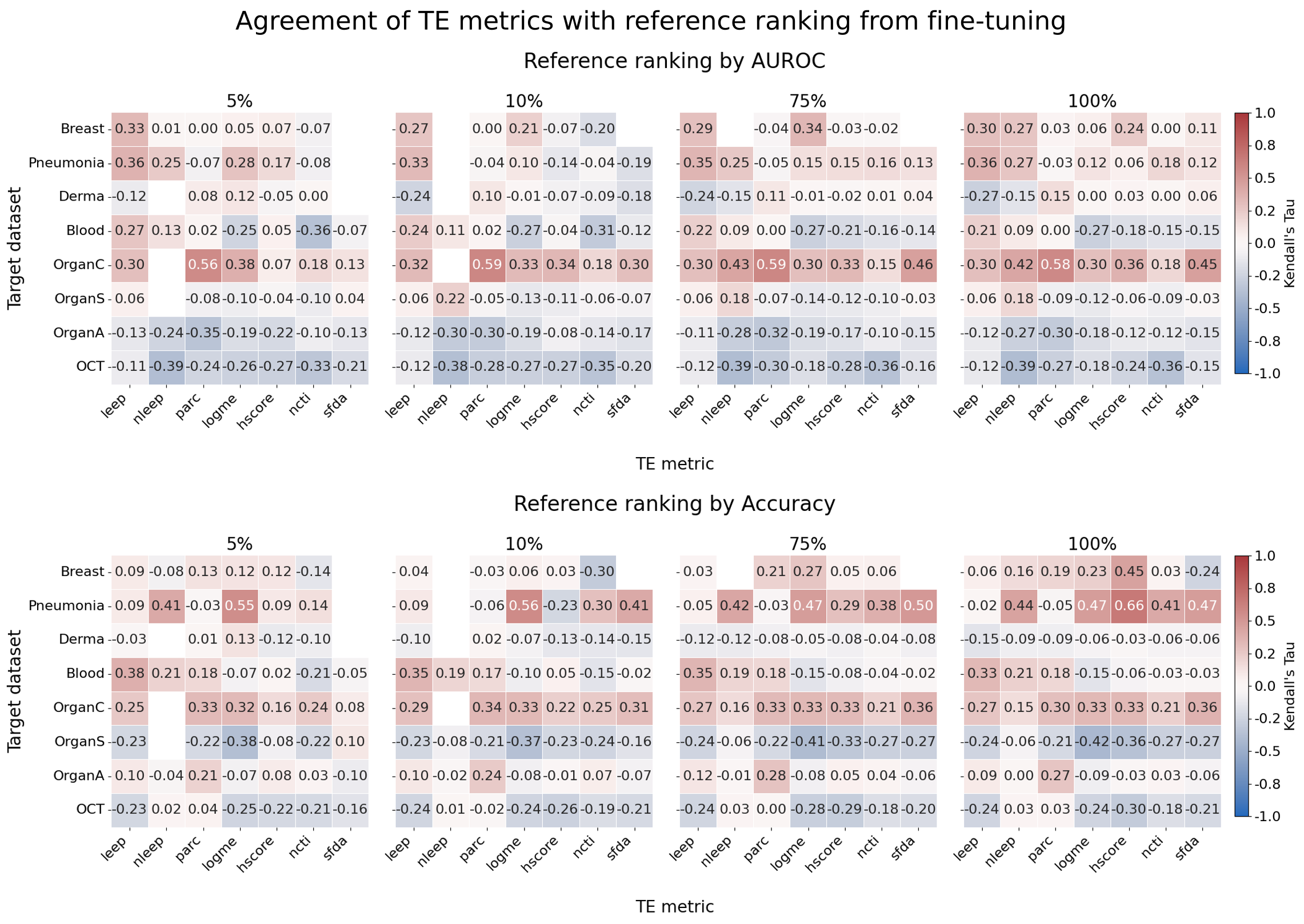}
    \caption{Average (over 5 random seeds for subsets < 100\%) pairwise Kendall's Tau ($stability_{ref}$) between TE metrics and reference ranking from fine-tuning, optimized separately for accuracy and AUROC. Blank fields indicate NaN values.}
    \label{fig:stability_ref_auc_acc_100}
\end{figure}

Since most TE metrics are designed to match the reference ranked by accuracy, we would expect this to be reflected in our experiments. However, as shown in Figure \ref{fig:stability_ref_auc_acc_100} for Kendall's Tau, the agreement between TE metrics and reference rankings remains low for both accuracy and AUROC across all target datasets. This is particularly interesting for the subset of 100\% as it represents the same data used for training during fine-tuning. This aligns with the work by Chaves et al. \cite{chaves_performance_2023}, who concluded that existing TE metrics do not reliably work for medical image classification tasks. Our findings clarify that this is not driven by the choice of evaluation metric, but rather a domain specific issue. Overall, we observe similar results from weighted Kendall's Tau and Spearman's rho which are shown in Appendix \ref{appx:stability_ref_from_finetuning}. 

\subsubsection{Reference rankings themselves are not robust:}

Next to showing that rankings are not the same across evaluation metrics, we also show that they are not robust across different subset sizes and random seeds. The results in terms of how many sources change ranks, are visually highly similar to Fig 7. For example, \textit{Blood} optimized for AUROC is ranked 3rd, 6th, or 10th depending on the representation of \textit{Breast} as target. For full results see appendix \ref{appx:reference_ranking_alignment_auc_acc_all}. Importantly, similar to TE metrics, these changes are not consistent across different target representations, meaning that a source model's position can improve under one target dataset representation while worsening under another. This further underscores the limited reliability of experiments that rely solely on a single subset. More broadly, this raises the question about whether comparing rankings with each other is meaningful. Adding to that, ranks discard the variation in absolute values. For instance, all source models achieve a AUROC of at least $0.99$ on \textit{Blood} using the full train data and are ranked based on only minor differences. At the same time, this can not easily be inferred from the TE metric scores.

Moreover, current TE metrics are designed in a way which assumes that adding a source will always result in a better performance compared to only training on the target dataset. There is no indication for cases where it might not be beneficial to apply TL due to potential negative transfer. For example, the ResNet-18 models from MedMNIST \cite{yang_medmnist_2023}, solely trained on either \textit{OrganA}, \textit{OrganC}, or \textit{OrganS} outperform all fine-tuned models in our work for these targets. Although the observed differences are minimal and the experimental setups vary, it nevertheless is a factor that should be considered.


\section{Conclusion}
\label{sec:discussion}

In this work, we studied the robustness of seven TE metrics with regards to different representation of the target dataset, as well as the impact of the evaluation metric used for reference ranking. We show that rankings from TE metrics vary among smaller target representations of the same size and class distribution but sampled using different random seeds. Importantly, the changes in rankings are not consistent across TE metrics or reference rankings, meaning that one source may improve for one metric while it does the opposite for another. Overall, this raises questions about the reliability of experimental setups where only one random seed is considered. Furthermore, we demonstrate that the evaluation metric used for the reference ranking influences the order of source models. While most TE metrics are designed to match accuracy, we observe a low agreement with the reference rankings across all target datasets for both accuracy and AUROC. 

Although this study provides valuable insights into the robustness of existing TE metrics, several limitations should be acknowledged. Most importantly, our generated miniature populations are based solely on class distributions and sample size which does not account for other clinically relevant factors, such as patient information. Furthermore, we only consider ResNet-18 as model architecture in our experiments. Therefore, our results are specific to this setup. Moreover, we base our work on MedMNIST, a collection of standardized biomedical images. However, this standardization makes it less close to real-world scenarios.

Building on the findings of this work, future research could explore how factors that influence the rankings of source models (such as evaluation metric) can be directly incorporated into TE metrics. Another worthwhile direction is to quantify the uncertainty of TE metric scores, offering a more reliable interpretation of rankings.

\section*{Acknowledgements}

This work is supported by Novo Nordisk Foundation grant NNF24OC00926. We also thank Inna Ermilova and Ties Robroek for their support with computational resources and training monitoring. Our thanks also goes to Beatrix Miranda Ginn Nielsen for help with the mathematical aspects of the problem definition.

%
%
\bibliographystyle{splncs04}
\bibliography{main}

\appendix
\newpage
\section{Fine-tuning results, GPU specifications, and energy consumption}

The following tables show the performance of the fine-tuned models on the test sets which the reference rankings discussed in section \ref{results_reference_ranking} are based on.

\label{appx:finetune_performance}
\renewcommand{\tablename}{A}
\setcounter{table}{0}


\begin{table}[!h]
\small
\centering
\caption{AUROC (rounded to three decimal places) on the test set after fine-tuning on 100\% of the targets train data, GPU specifications, and energy consumption (rounded to two decimal places) in kilowatt-hours (kWh). The best-performing source model is shown in bold.}
\label{tab:results_finetuning_100pct_auc}

\begin{tabularx}{\linewidth}{l|c|c|c|c|c|c|c|c}
\toprule

\diagbox[width=1.6cm,height=1cm]{\textbf{Source}}{\textbf{Target}} &
\textbf{Breast} &
\textbf{Pneumonia} &
\textbf{Derma} &
\textbf{Blood} &
\textbf{OrganC} &
\textbf{OrganS} &
\textbf{OrganA} &
\textbf{OCT} \\

\hline

ImageNet & 0.879 & \textbf{0.993} & \textbf{0.981} & \textbf{0.999} & \textbf{0.990} & 0.964 & 0.983 & 0.992 \\
Blood & 0.895 & 0.985 & 0.935 & - & 0.984 & 0.965 & 0.989 & 0.996 \\
Breast & - & 0.985 & 0.928 & 0.998 & 0.985  & 0.967 & 0.988 & 0.995 \\
Derma & 0.872 & 0.983 & - & 0.997 & 0.984 & 0.960 & 0.990 & 0.997 \\
OCT & 0.885 & 0.988 & 0.938 & 0.998 & 0.979 & 0.964 & 0.988 & - \\
OrganA & \textbf{0.916} & 0.990 & 0.942 & 0.998 & 0.989 & 0.969 & - & 0.997\\
OrganC & 0.873 & 0.987 & 0.937 & 0.998 & - & 0.965 & 0.987 & 0.996 \\
OrganS & 0.892 & 0.983 & 0.939 & 0.998 & \textbf{0.990} & - & 0.987 & \textbf{0.998} \\
Path & 0.840 & 0.988 & 0.955 & \textbf{0.999} & 0.982 & 0.968 & 0.979 & \textbf{0.998}\\
Pneumonia & 0.854 & - & 0.933 & 0.998 & 0.985 & 0.964 & 0.989 & \textbf{0.998}\\
Retina & 0.839 & 0.968 & 0.925 & 0.998 & 0.977 & 0.963 & \textbf{0.992} & \textbf{0.998}\\
Tissue & 0.897 & 0.986 & 0.946 & 0.997 & 0.984 & 0.963 & 0.979 & 0.996\\
Chest & 0.876 & 0.984 & 0.938 & 0.998 & 0.977 & \textbf{0.970} & 0.990 & 0.997\\

\midrule

GPU & GB10 & GB10 & GB10 & GB10 & A100 & GB10 & A100 & A100 \\

kWh & 0.42 & 2.61 & 4.06 & 8.42 & 13.01 & 11.14 & 33.44 & 65.68 \\

\bottomrule
\end{tabularx}
\end{table}

\begin{table}[!h]
\small
\centering
\caption{Accuracy (rounded to three decimal places) on the test set after fine-tuning on 100\% of the targets train data, GPU specifications, and energy consumption (rounded to two decimal places) in kilowatt-hours (kWh). The best-performing source model is shown in bold.}
\label{tab:results_finetuning_100pct_acc}

\begin{tabularx}{\linewidth}{l|c|c|c|c|c|c|c|c}
\toprule

\diagbox[width=1.6cm,height=1cm]{\textbf{Source}}{\textbf{Target}} &
\textbf{Breast} &
\textbf{Pneumonia} &
\textbf{Derma} &
\textbf{Blood} &
\textbf{OrganC} &
\textbf{OrganS} &
\textbf{OrganA} &
\textbf{OCT} \\

\hline

ImageNet & \textbf{0.885} & \textbf{0.957} & \textbf{0.874} & \textbf{0.974} & 0.821 & 0.721 & 0.810 & 0.867\\
Blood & 0.859 & 0.941 & 0.772 & - & 0.814 & 0.703 & 0.846 & 0.879\\
Breast & - & 0.885 & 0.772 & 0.912 & 0.834 & 0.716 & 0.849 & 0.871\\
Derma & 0.776 & 0.921 & - & 0.904 & 0.804 & 0.684 & 0.865 & 0.886\\
OCT & 0.859 & 0.872 & 0.783 & 0.951 & 0.795 & 0.709 & 0.833 &-\\
OrganA & 0.859 & 0.950 & 0.782 & 0.929 & 0.842 & 0.699 & - & 0.873\\
OrganC & 0.814 & 0.921 & 0.791 & 0.953 & - & 0.707 & 0.809 & 0.874\\
OrganS & 0.833 & 0.837 & 0.778 & 0.946 & \textbf{0.852} & - & 0.871 & 0.875\\
Path & 0.859 & 0.833 & 0.809 & 0.962 & 0.814 & 0.731 & \textbf{0.897} & \textbf{0.905}\\
Pneumonia & 0.821 & - & 0.766 & 0.955 & 0.838 & \textbf{0.745} & 0.785 & 0.863\\
Retina & 0.808 & 0.878 & 0.768 & 0.916 & 0.821 & 0.718 & 0.851 & 0.854\\
Tissue & 0.833 & 0.899 & 0.781 & 0.950 & 0.803 & 0.738 & 0.849 & 0.866\\
Chest & 0.821 & 0.888 & 0.776 & 0.915 & 0.775 & 0.741 & 0.853 & 0.893\\

\midrule

GPU & GB10 & GB10 & GB10 & GB10 & A100 & GB10 & A100 & A100 \\

kWh & 0.46 & 3.56 & 3.65 & 7.03 & 12.64 & 9.38 & 28.60 & 68.88\\

\bottomrule
\end{tabularx}
\end{table}


\begin{table}[!h]
\small
\centering
\caption{AUROC (rounded to three decimal places) on the test set after fine-tuning on 5\% of the targets train data sampled with a random seed of 42, GPU specifications, and energy consumption (rounded to two decimal places) in kilowatt-hours (kWh). The best-performing source model is shown in bold.}
\label{tab:results_finetuning_5pct_split1_auc}

\begin{tabularx}{\linewidth}{l|c|c|c|c|c|c|c|c}
\toprule

\diagbox[width=1.6cm,height=1cm]{\textbf{Source}}{\textbf{Target}} &
\textbf{Breast} &
\textbf{Pneumonia} &
\textbf{Derma} &
\textbf{Blood} &
\textbf{OrganC} &
\textbf{OrganS} &
\textbf{OrganA} &
\textbf{OCT} \\

\hline

ImageNet & 0.681 & \textbf{0.984} & \textbf{0.867} & \textbf{0.997} & 0.960 & \textbf{0.950} & \textbf{0.983} & \textbf{0.993}\\
Blood & 0.744 & 0.934 & 0.832 & - & 0.940 & 0.931 & 0.972 & 0.977\\
Breast & - & 0.947 & 0.811 & 0.973 & 0.944 & 0.911 & 0.982 & 0.954\\
Derma & 0.786 & 0.946 & - & 0.978 & 0.958 & 0.934 & 0.967 & 0.978 \\
OCT & 0.655 & 0.950 & 0.834 & 0.972 & 0.954 & 0.924 & 0.963 & - \\
OrganA & 0.792 & 0.942 & 0.826 & 0.985 & \textbf{0.981} & 0.942 & - & 0.984\\
OrganC & 0.768 & 0.935 & 0.843 & 0.985 & - & 0.940 & 0.980 & 0.976\\
OrganS & 0.758 & 0.931 & 0.838 & 0.982 & 0.956 & - & 0.976 & 0.982\\
Path & 0.737 & 0.916 & 0.842 & 0.991 & 0.952 & 0.937 & 0.969 & 0.978\\
Pneumonia & \textbf{0.793} & - & 0.832 & 0.970 & 0.960 & 0.917 & 0.969 & 0.983\\
Retina & 0.706 & 0.913 & 0.805 & 0.976 & 0.941 & 0.923 & 0.971 & 0.971\\
Tissue & 0.738 & 0.942 & 0.833 & 0.974 & 0.964 & 0.935 & 0.970 & 0.985\\
Chest & 0.684 & 0.970 & 0.853 & 0.977 & 0.956 & 0.935 & 0.973 & 0.985\\

\midrule

GPU & GB10 & GB10 & GB10 & GB10 & GB10 & GB10 & GB10 & GB10\\

kWh & 0.12 & 0.28 & 0.40 & 0.63 & 0.96 & 0.93 & 2.17 & 4.53\\





\bottomrule
\end{tabularx}
\end{table}

\begin{table}[!h]
\small
\centering
\caption{Accuracy (rounded to three decimal places) on the test set after fine-tuning on 5\% of the targets train data sampled with a random seed of 42, GPU specifications, and energy consumption (rounded to two decimal places) in kilowatt-hours (kWh). The best-performing source model is shown in bold.}
\label{tab:results_finetuning_5pct_split1_acc}

\begin{tabularx}{\linewidth}{l|c|c|c|c|c|c|c|c}
\toprule

\diagbox[width=1.6cm,height=1cm]{\textbf{Source}}{\textbf{Target}} &
\textbf{Breast} &
\textbf{Pneumonia} &
\textbf{Derma} &
\textbf{Blood} &
\textbf{OrganC} &
\textbf{OrganS} &
\textbf{OrganA} &
\textbf{OCT} \\

\hline

ImageNet & \textbf{0.795} & \textbf{0.941} & \textbf{0.753} & \textbf{0.920} & 0.722 & \textbf{0.665} & \textbf{0.818} & 0.833\\
Blood & 0.731 & 0.806 & 0.700 & - & 0.675 & 0.605 & 0.763 & 0.732\\
Breast & - & 0.752 & 0.677 & 0.751 & 0.648 & 0.566 & 0.741 & 0.658 \\
Derma & 0.705 & 0.841 & - & 0.818 & 0.686 & 0.588 & 0.716 & 0.763 \\
OCT & 0.724 & 0.846 & 0.693 & 0.738 & 0.628 & 0.614 & 0.703 & - \\
OrganA & 0.731 & 0.795 & 0.706 & 0.828 & \textbf{0.779} & 0.610 & - & 0.743 \\
OrganC & 0.750 & 0.830 & 0.705 & 0.837 & - & 0.608 & 0.813 & 0.725 \\
OrganS & 0.782 & 0.817 & 0.716 & 0.846 & 0.706 & - & 0.747 & 0.807 \\
Path & 0.750 & 0.832 & 0.703 & 0.878 & 0.589 & 0.575 & 0.784 & 0.748 \\
Pneumonia & 0.724 & - & 0.701 & 0.794 & 0.666 & 0.511 & 0.694 & 0.732\\
Retina & 0.744 & 0.833 & 0.674 & 0.807 & 0.628 & 0.515 & 0.732 & 0.688\\
Tissue & 0.769 & 0.870 & 0.707 & 0.800 & 0.678 & 0.637 & 0.808 & \textbf{0.853}\\
Chest & 0.744 & 0.873 & 0.706 & 0.745 & 0.699 & 0.593 & 0.738 & 0.750 \\

\midrule

GPU & GB10 & GB10 & GB10 & GB10 & GB10 & GB10 & GB10 & GB10\\

kWh & 0.24 & 0.35 & 0.68 & 0.60 & 0.86 & 0.93 & 2.12 & 5.05 \\

\bottomrule
\end{tabularx}
\end{table}


\begin{table}[!h]
\small
\centering
\caption{AUROC (rounded to three decimal places) on the test set after fine-tuning on 5\% of the targets train data sampled with a random seed of 43, GPU specifications, and energy consumption (rounded to two decimal places) in kilowatt-hours (kWh). The best-performing source model is shown in bold. A100* refers to NVIDIA A100 80GB.}
\label{tab:results_finetuning_5pct_split2_auc}

\begin{tabularx}{\linewidth}{l|c|c|c|c|c|c|c|c}
\toprule

\diagbox[width=1.6cm,height=1cm]{\textbf{Source}}{\textbf{Target}} &
\textbf{Breast} &
\textbf{Pneumonia} &
\textbf{Derma} &
\textbf{Blood} &
\textbf{OrganC} &
\textbf{OrganS} &
\textbf{OrganA} &
\textbf{OCT} \\

\hline

ImageNet & 0.793 & 0.971 &\textbf{ 0.866} & \textbf{0.996} & 0.961 & 0.942 & 0.974 & \textbf{0.994}\\
Blood & 0.759 & 0.949 & 0.836 & - & 0.946 & 0.937 & 0.960 & 0.964 \\
Breast & - & 0.952 & 0.821 & 0.983 & 0.922 & 0.928 & 0.967 & 0.965\\
Derma & 0.793 & 0.943 & - & 0.980 & 0.951 & 0.940 & 0.970 & 0.960 \\
OCT & 0.779 & 0.956 & 0.833 & 0.956 & 0.945 & 0.946 & 0.971 & - \\
OrganA & 0.817 & 0.951 & 0.843 & 0.986 & \textbf{0.979} & 0.944 & - & 0.978 \\
OrganC & 0.753 & 0.944 & 0.843 & 0.985 & - & \textbf{0.949} & 0.979 & 0.979\\
OrganS & 0.769 & 0.950 & 0.849 & 0.987 & 0.962 & - & 0.972 & 0.965\\
Path & 0.747 & 0.945 & 0.864 & 0.992 & 0.926 & 0.939 & 0.954 & 0.972\\
Pneumonia & 0.792 & - & 0.834 & 0.978 & 0.953 & 0.925 & 0.973 & 0.971\\
Retina & 0.772 & 0.938 & 0.830 & 0.980 & 0.926 & 0.907 & 0.967 & 0.975\\
Tissue & \textbf{0.821} & 0.945 & 0.838 & 0.978 & 0.955 & 0.946 & 0.980 & 0.978\\
Chest & 0.799 & \textbf{0.975} & 0.839 & 0.970 & 0.951 & 0.943 & \textbf{0.985} & 0.980\\

\midrule

GPU & A100* & A100* & A100* & A100* & A100* & A100* & A100* & A100*\\

kWh & 0.37 & 0.59 & 0.95 & 0.96 & 1.7 & 1.76 & 3.69 & 6.12\\





\bottomrule
\end{tabularx}
\end{table}

\begin{table}[!h]
\small
\centering
\caption{Accuracy (rounded to three decimal places) on the test set after fine-tuning on 5\% of the targets train data sampled with a random seed of 43, GPU specifications, and energy consumption (rounded to two decimal places) in kilowatt-hours (kWh). The best-performing source model is shown in bold.}
\label{tab:results_finetuning_5pct_split2_acc}

\begin{tabularx}{\linewidth}{l|c|c|c|c|c|c|c|c}
\toprule

\diagbox[width=1.6cm,height=1cm]{\textbf{Source}}{\textbf{Target}} &
\textbf{Breast} &
\textbf{Pneumonia} &
\textbf{Derma} &
\textbf{Blood} &
\textbf{OrganC} &
\textbf{OrganS} &
\textbf{OrganA} &
\textbf{OCT} \\

\hline

ImageNet & \textbf{0.814} & 0.865 & \textbf{0.741} & \textbf{0.940} & \textbf{0.719} & 0.616 & 0.782 & \textbf{0.858}\\
Blood & 0.782 & 0.782 & 0.676 & - & 0.591 & 0.635 & 0.736 & 0.735 \\
Breast & - & 0.771 & 0.686 & 0.842 & 0.623 & 0.530 & 0.739 & 0.676 \\
Derma & 0.776 & 0.857 & - & 0.841 & 0.619 & 0.610 & 0.709 & 0.726\\
OCT & 0.763 & 0.848 & 0.688 & 0.581 & 0.630 & 0.627 & 0.730 & - \\
OrganA & 0.788 & 0.816 & 0.705 & 0.834 & 0.717 & 0.634 & - & 0.738\\
OrganC & 0.788 & 0.848 & 0.697 & 0.831 & - & 0.635 & 0.758 & 0.744\\
OrganS & 0.769 & 0.867 & 0.702 & 0.857 & 0.712 & - & 0.780 & 0.756\\
Path & 0.750 & 0.846 & 0.687 & 0.886 & 0.660 & 0.632 & 0.719 & 0.757\\
Pneumonia & 0.788 & - & 0.694 & 0.740 & 0.654 & 0.597 & 0.665 & 0.756 \\
Retina & 0.744 & 0.819 & 0.672 & 0.825 & 0.610 & 0.597 & 0.726 & 0.646 \\
Tissue & 0.750 & \textbf{0.897} & 0.708 & 0.828 & 0.685 & \textbf{0.658} & \textbf{0.793} & 0.774\\
Chest & 0.776 & 0.865 & 0.694 & 0.743 & 0.671 & 0.614 & 0.701 & 0.787 \\

\midrule

GPU & A30 & A100 & A30 & A30 & GB10 & A100 & A30 & A30\\

kWh & 0.33 & 0.54 & 1.12 & 0.84 & 0.85 & 1.32 & 3.69 & 6.03 \\

\bottomrule
\end{tabularx}
\end{table}

\clearpage
\section{Intra-metric stability for Kendall's Tau, weighted Kendall's Tau, and Spearman's rho}
\label{appx:intra_metric_stability}
\renewcommand{\figurename}{B}
\setcounter{figure}{0}

The following figure shows intra-metric stability results for all three correlation coefficients which are discussed in section \ref{results_intra_stability}.

\begin{figure}[h]
    \centering
    \includegraphics[width=1\textwidth]{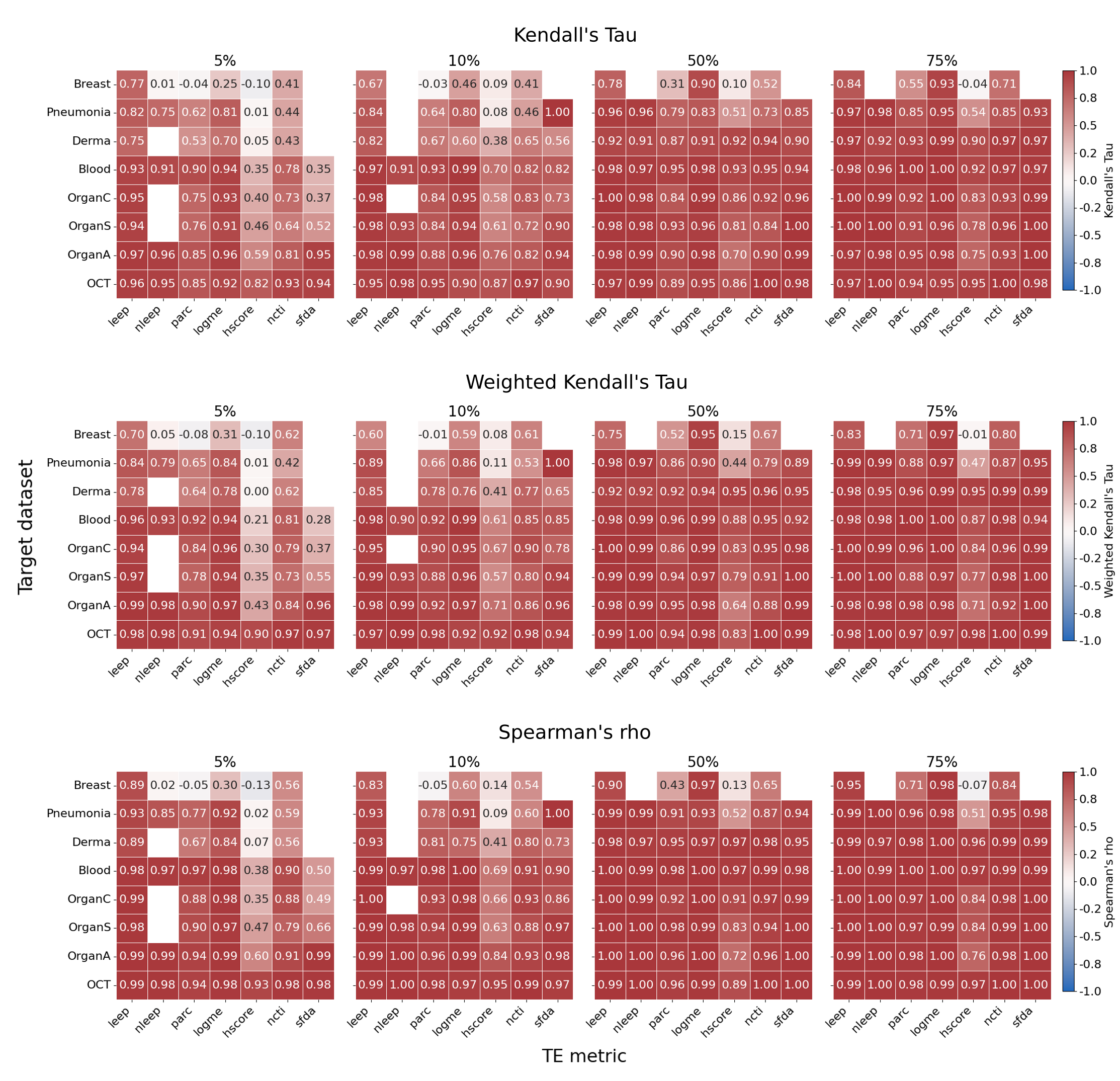} 
    \caption{Average (over 5 random seeds) pairwise correlation ($stability_{intra}$) across target subsets of varying sizes for different TE metrics. The target datasets are sorted by absolute train set size in ascending order. Blank fields indicate NaN values.}
    \label{fig:stability_intra_all}
\end{figure}

\clearpage
\section{Intra-metric stability for modified $\mathcal{N}LEEP$}
\label{appx:intra_metric_stability_modified_nleep}
\renewcommand{\figurename}{C}
\setcounter{figure}{0}

The following figure shows intra-metric stability results for a modified version of $\mathcal{N}LEEP$ as described in section \ref{results_intra_stability}.

\begin{figure}[h]
    \centering
    \includegraphics[width=1\textwidth]{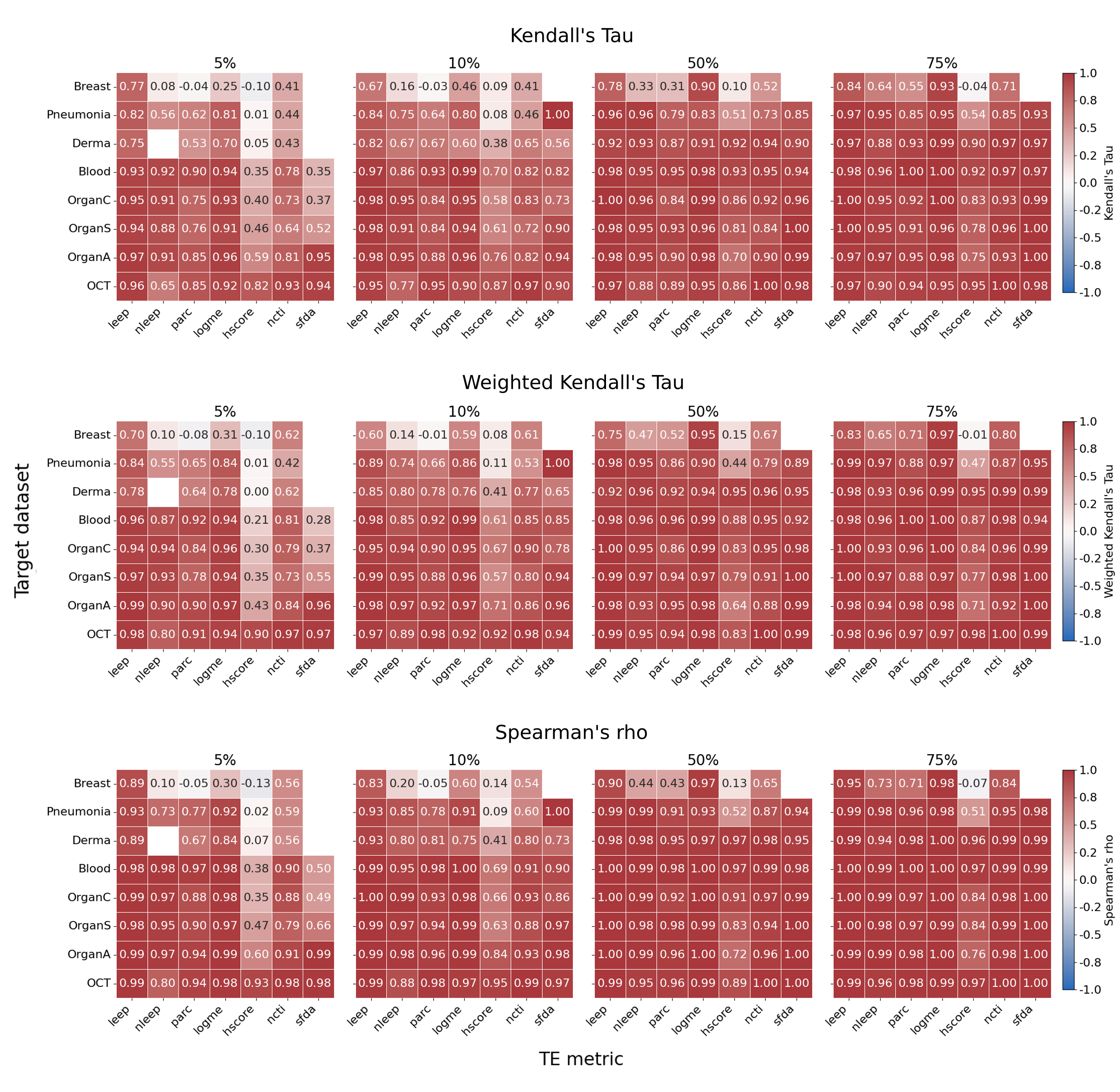} 
    \caption{Average (over 5 random seeds) pairwise correlation ($stability_{intra}$) across target subsets of varying sizes for different TE metrics. The target datasets are sorted by absolute train set size in ascending order. Blank fields indicate NaN values. $\mathcal{N}LEEP$ adjusted by setting the principal component analysis energy to $0.9$ (from $0.8$) and the number of Gaussian components to $1$ (from $5$) per class.}
    \label{fig:stability_intra_all_nleep_modified}
\end{figure}

\clearpage
\section{Ranking stability between subsets and reference population for Kendall's Tau, weighted Kendall's Tau, and Spearman's rho}
\label{appx:stability_ref_from_metric_all}
\renewcommand{\figurename}{D}
\setcounter{figure}{0}

The following figure presents additional results of TE metrics for the agreement between subsets and the full target dataset as reference, as discussed in section \ref{results_intra_stability}.

\begin{figure}[h]
    \centering
    \includegraphics[width=1\textwidth]{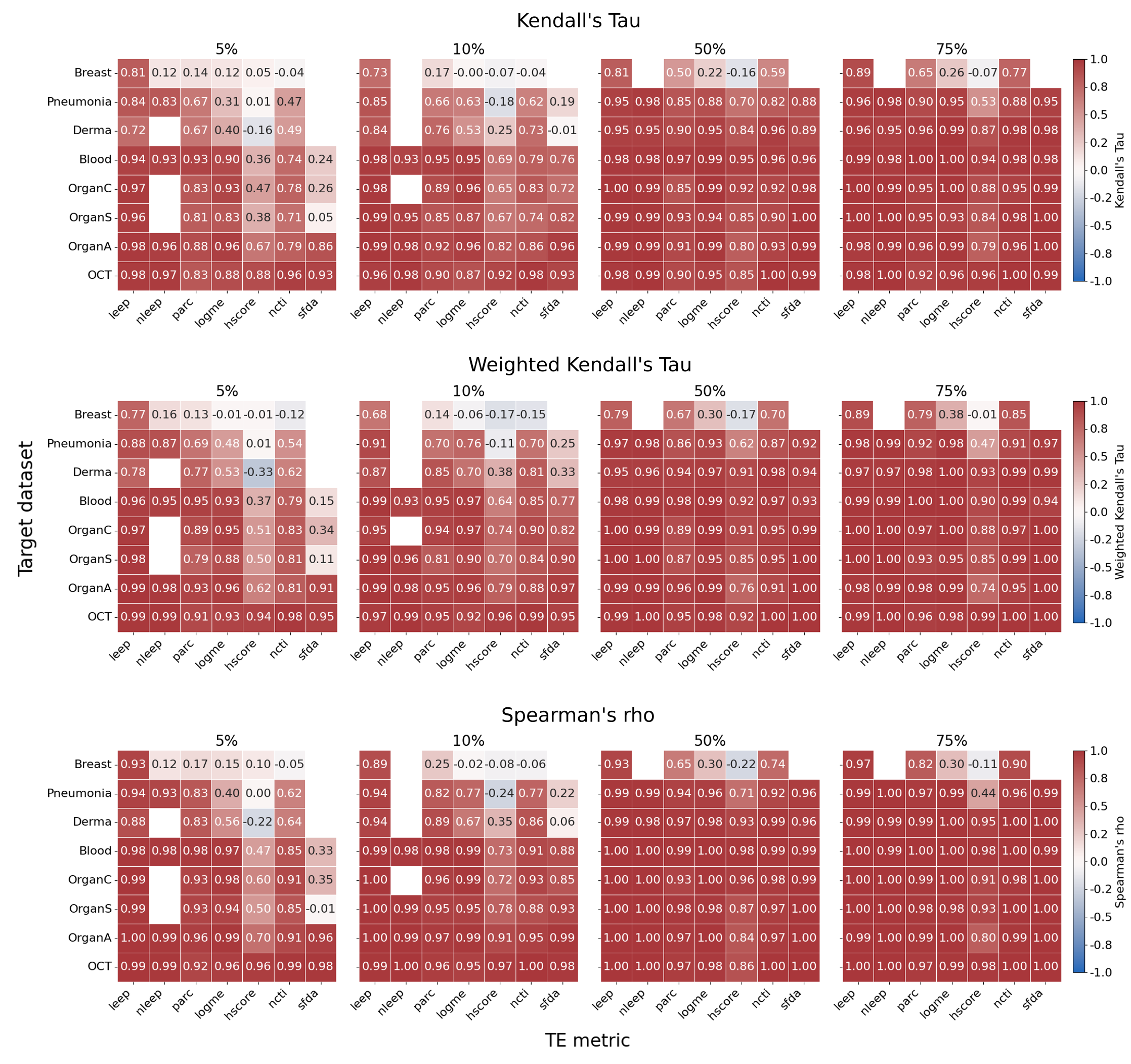} 
    \caption{Average (over 5 random seeds) pairwise correlation ($stability_{ref}$) between rankings obtained from target subsets of varying fraction sizes and the ranking obtained using 100\% of the data. The target datasets are sorted by absolute train set size in ascending order. Blank fields indicate NaN values.}
    \label{fig:stability_ref_from_metric_all}
\end{figure}

\clearpage
\section{Rank agreement between metrics}
\label{appx:stability_inter}
\renewcommand{\figurename}{E}
\setcounter{figure}{0}

The following figures show additional results for the rank agreement between TE metrics which is covered in section \ref{results_intra_stability}.

\begin{figure}[!h]
    \centering
    \includegraphics[width=1\textwidth]{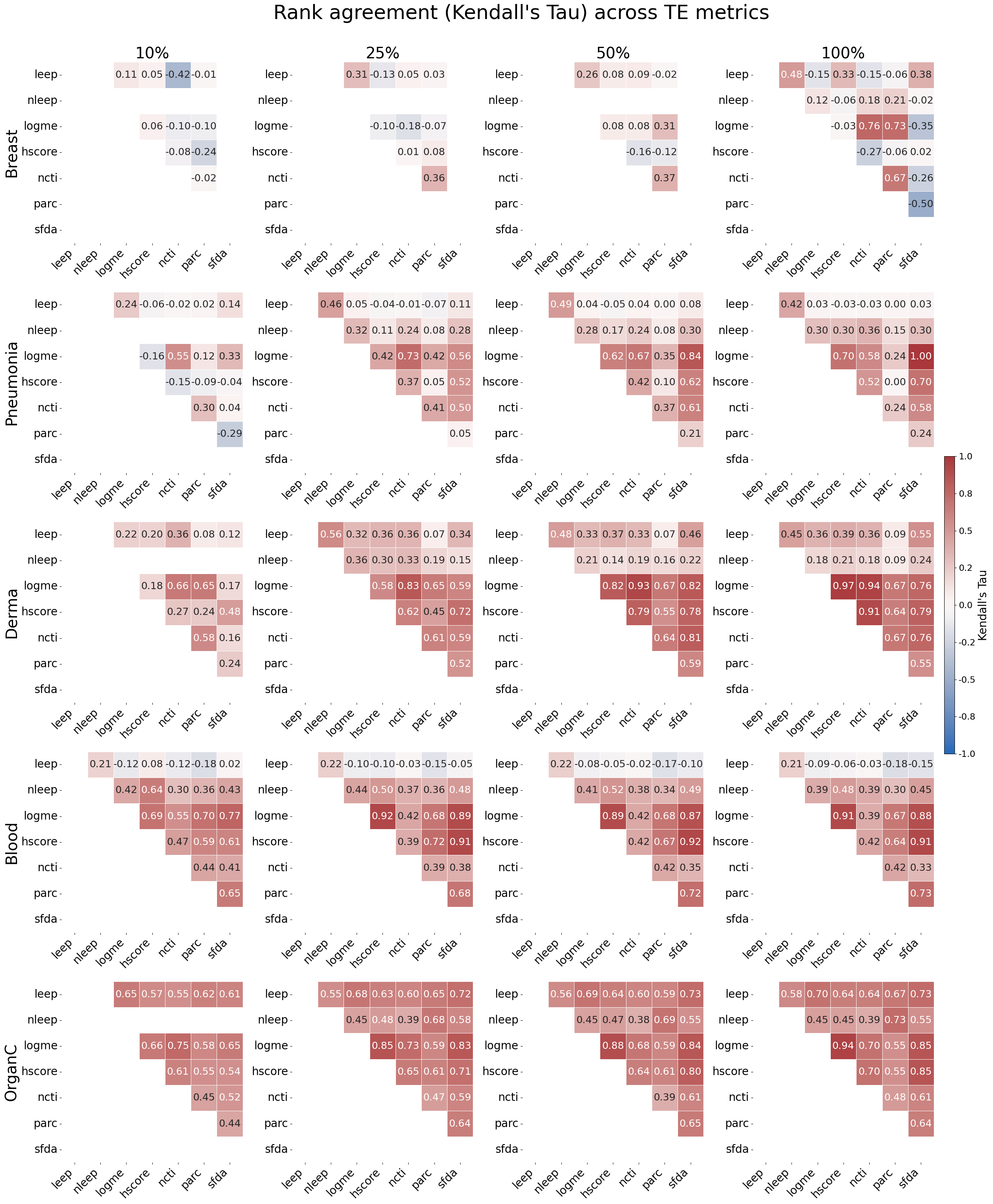} 
    \caption{Average pairwise Kendall's Tau ($stability_{inter}$) between rankings from different TE metrics. For subset sizes $< 100\%$, agreement is computed between subsets generated with identical random seeds and then averaged across 5 seeds.}
    \label{fig:stability_inter_kendalls_first_half}
\end{figure}

\begin{figure}[h]
    \centering
    \includegraphics[width=1\textwidth]{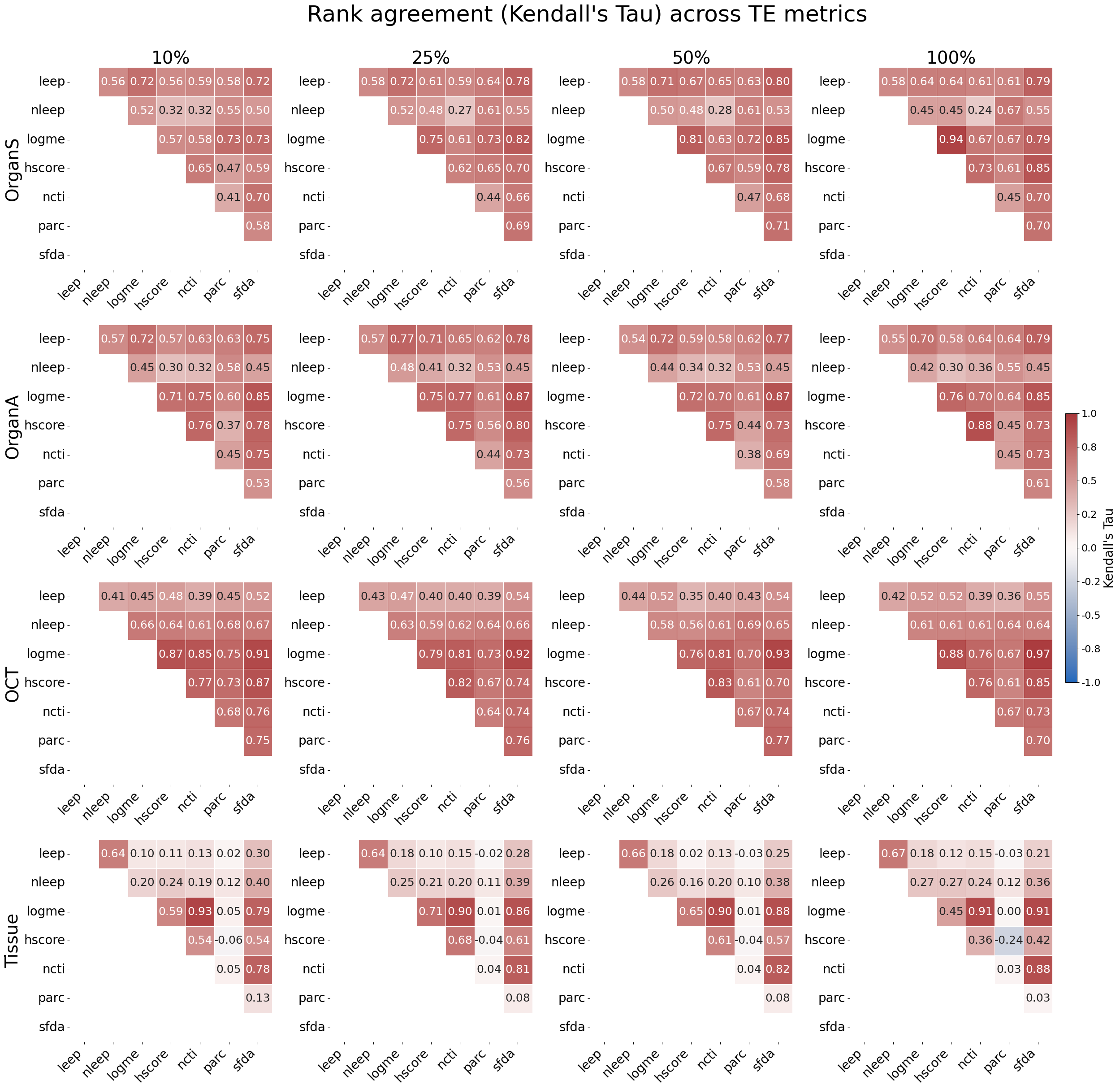} 
    \caption{Average pairwise Kendall's Tau ($stability_{inter}$) between rankings from different TE metrics. For subset sizes $< 100\%$, agreement is computed between subsets generated with identical random seeds and then averaged across 5 seeds.}
    \label{fig:stability_inter_kendalls_second_half}
\end{figure}

\begin{figure}[h]
    \centering
    \includegraphics[width=1\textwidth]{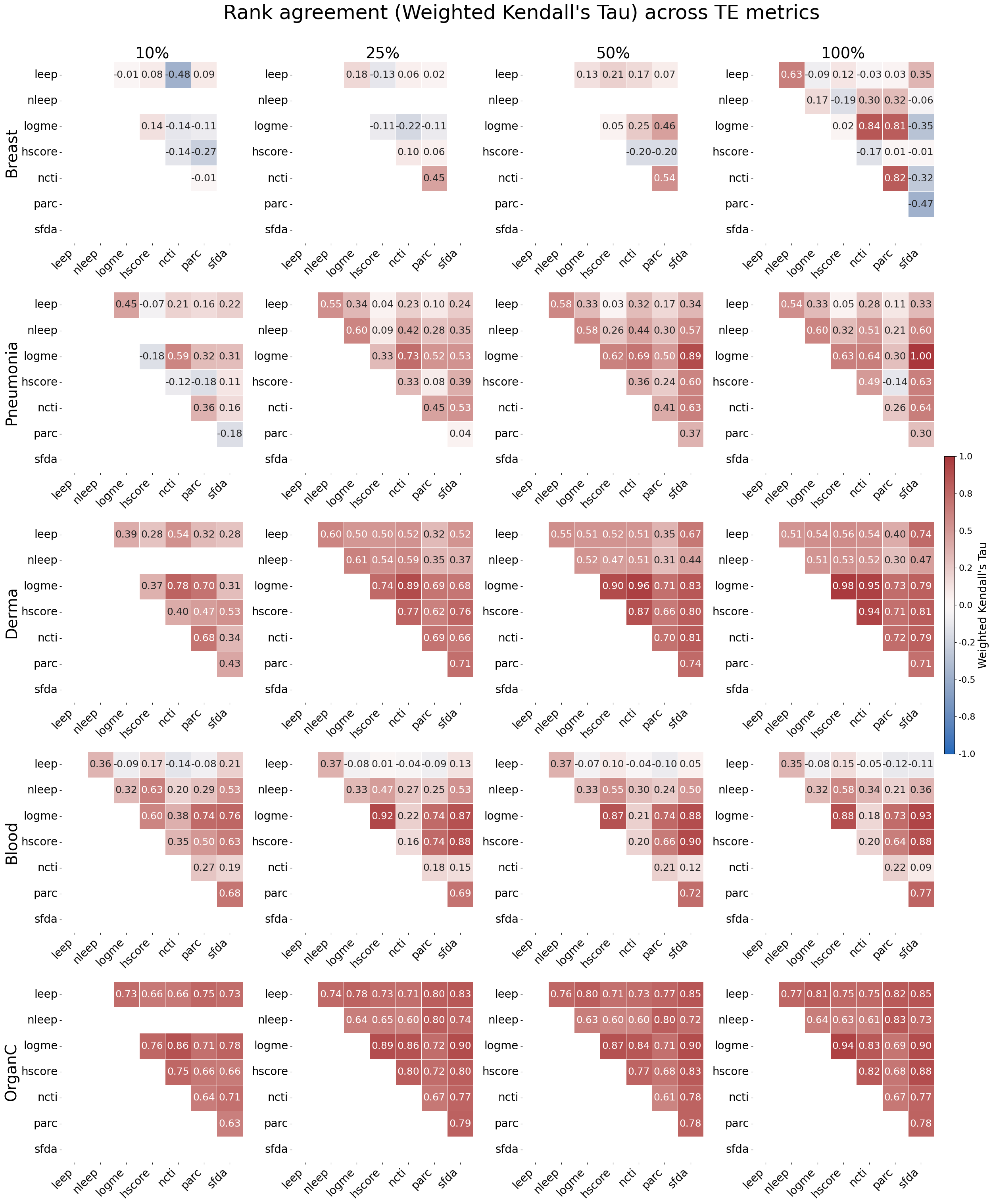} 
    \caption{Average pairwise weighted Kendall's Tau ($stability_{inter}$) between rankings from different TE metrics. For subset sizes $< 100\%$, agreement is computed between subsets generated with identical random seeds and then averaged across 5 seeds.}
    \label{fig:stability_inter_weighted_kendalls_first_half}
\end{figure}

\begin{figure}[h]
    \centering
    \includegraphics[width=1\textwidth]{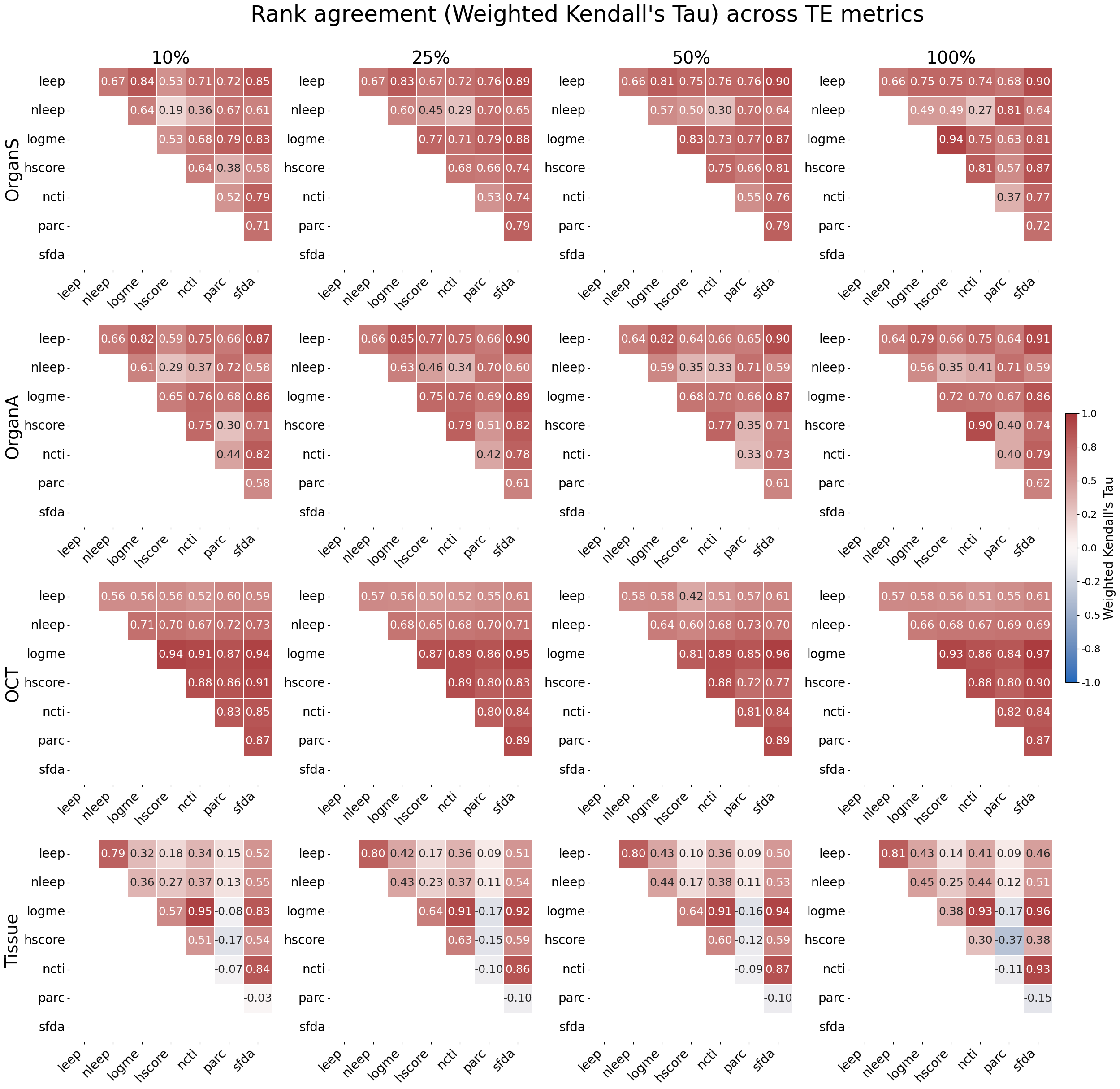} 
    \caption{Average pairwise weighted Kendall's Tau ($stability_{inter}$) between rankings from different TE metrics. For subset sizes $< 100\%$, agreement is computed between subsets generated with identical random seeds and then averaged across 5 seeds.}
    \label{fig:stability_inter_weighted_kendalls_second_half}
\end{figure}

\begin{figure}[h]
    \centering
    \includegraphics[width=1\textwidth]{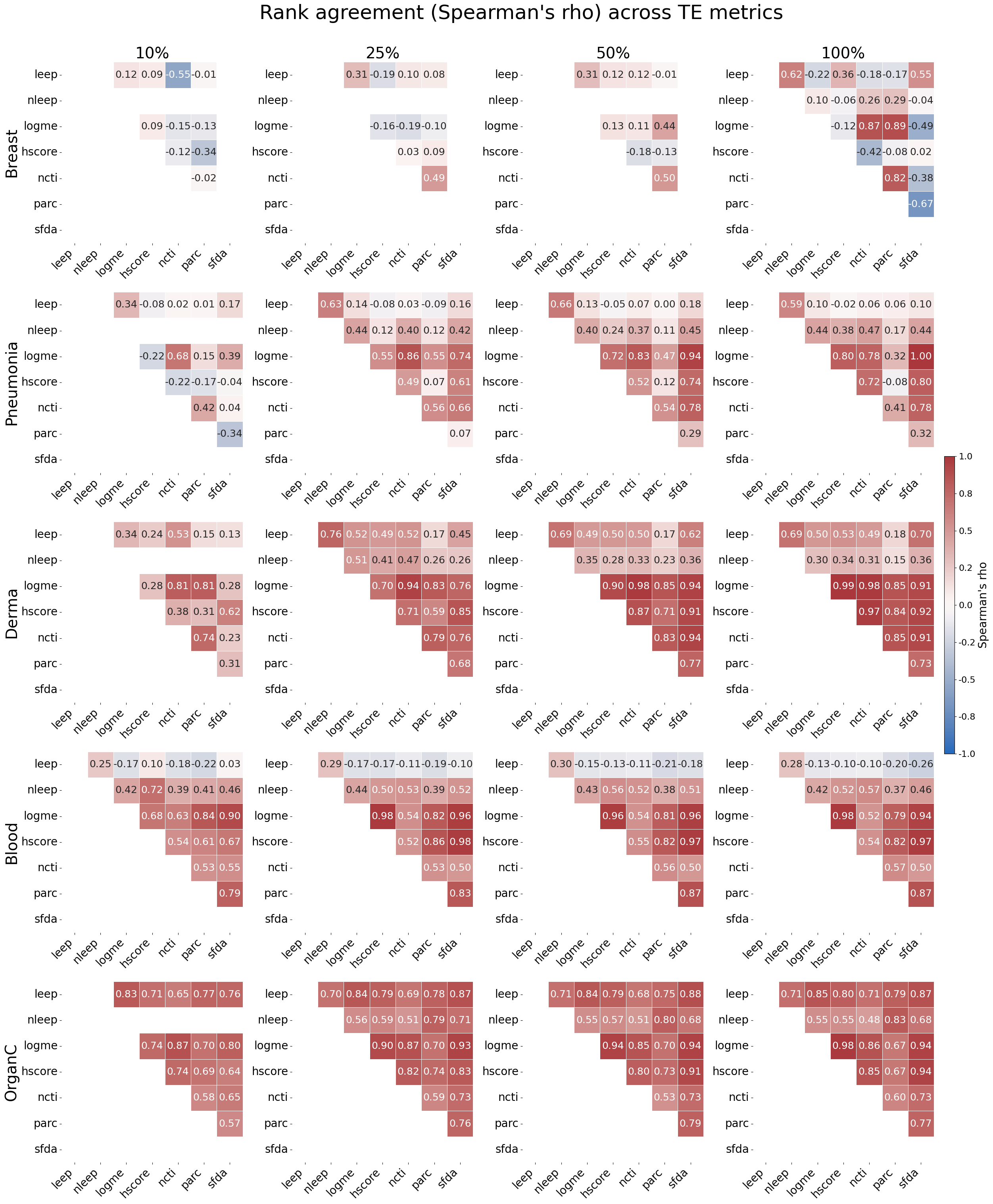} 
    \caption{Average pairwise Spearman's rho ($stability_{inter}$) between rankings from different TE metrics. For subset sizes $< 100\%$, agreement is computed between subsets generated with identical random seeds and then averaged across 5 seeds.}
    \label{fig:stability_inter_speraman_first_half}
\end{figure}

\begin{figure}[h]
    \centering
    \includegraphics[width=1\textwidth]{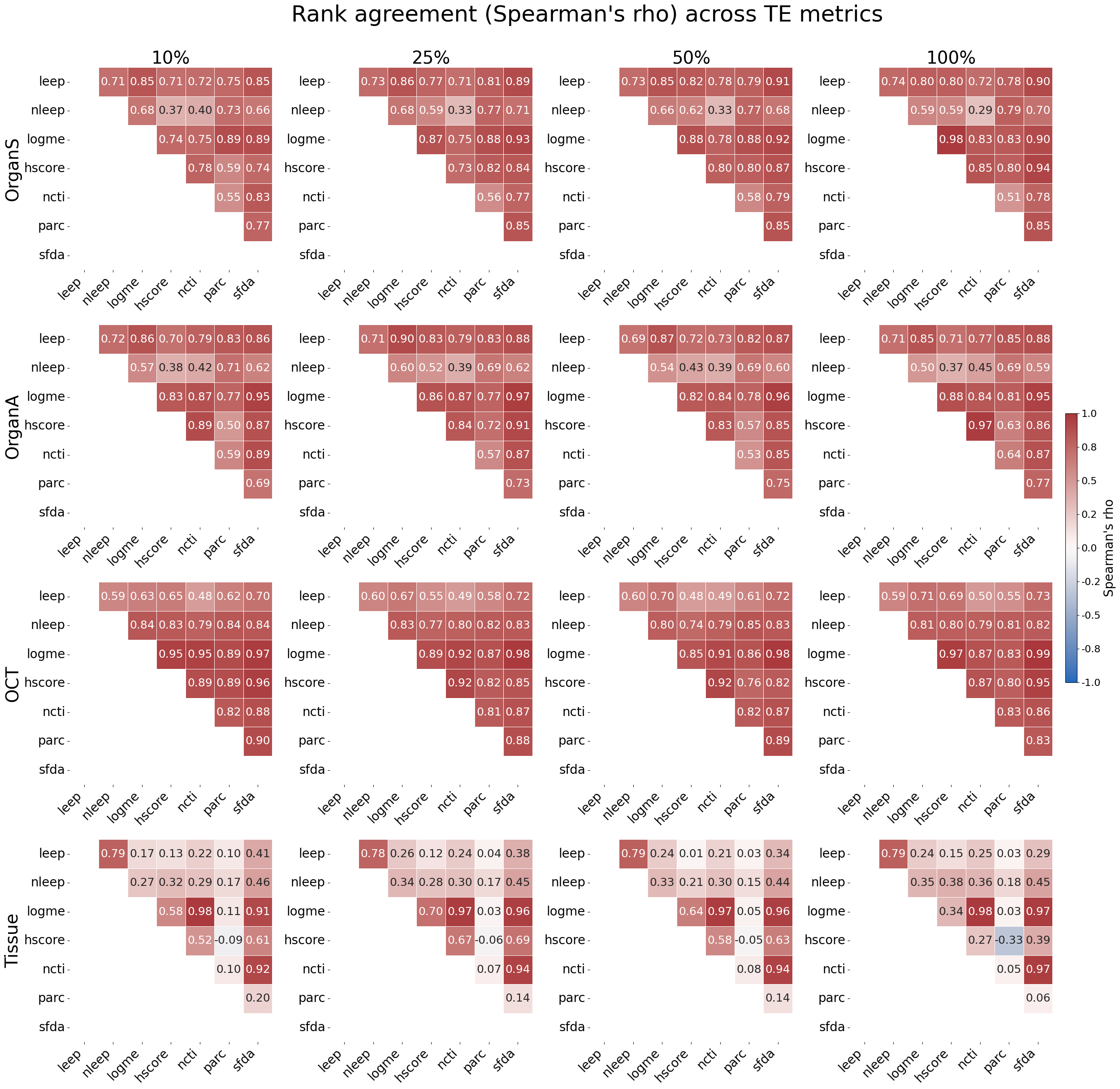} 
    \caption{Average pairwise Spearman's rho ($stability_{inter}$) between rankings from different TE metrics. For subset sizes $< 100\%$, agreement is computed between subsets generated with identical random seeds and then averaged across 5 seeds.}
    \label{fig:stability_inter_spearman_second_half}
\end{figure}

\clearpage
\section{Rank agreement between reference rankings seperatly optimized for accuracy and AUROC}
\label{appx:reference_ranking_alignment_auc_acc_all}
\renewcommand{\figurename}{F}
\setcounter{figure}{0}

The following figures show additional results for the limited robustness of reference rankings, as discussed in section \ref{results_reference_ranking}.

\begin{figure}[h!]
    \centering
    \includegraphics[width=1\textwidth]{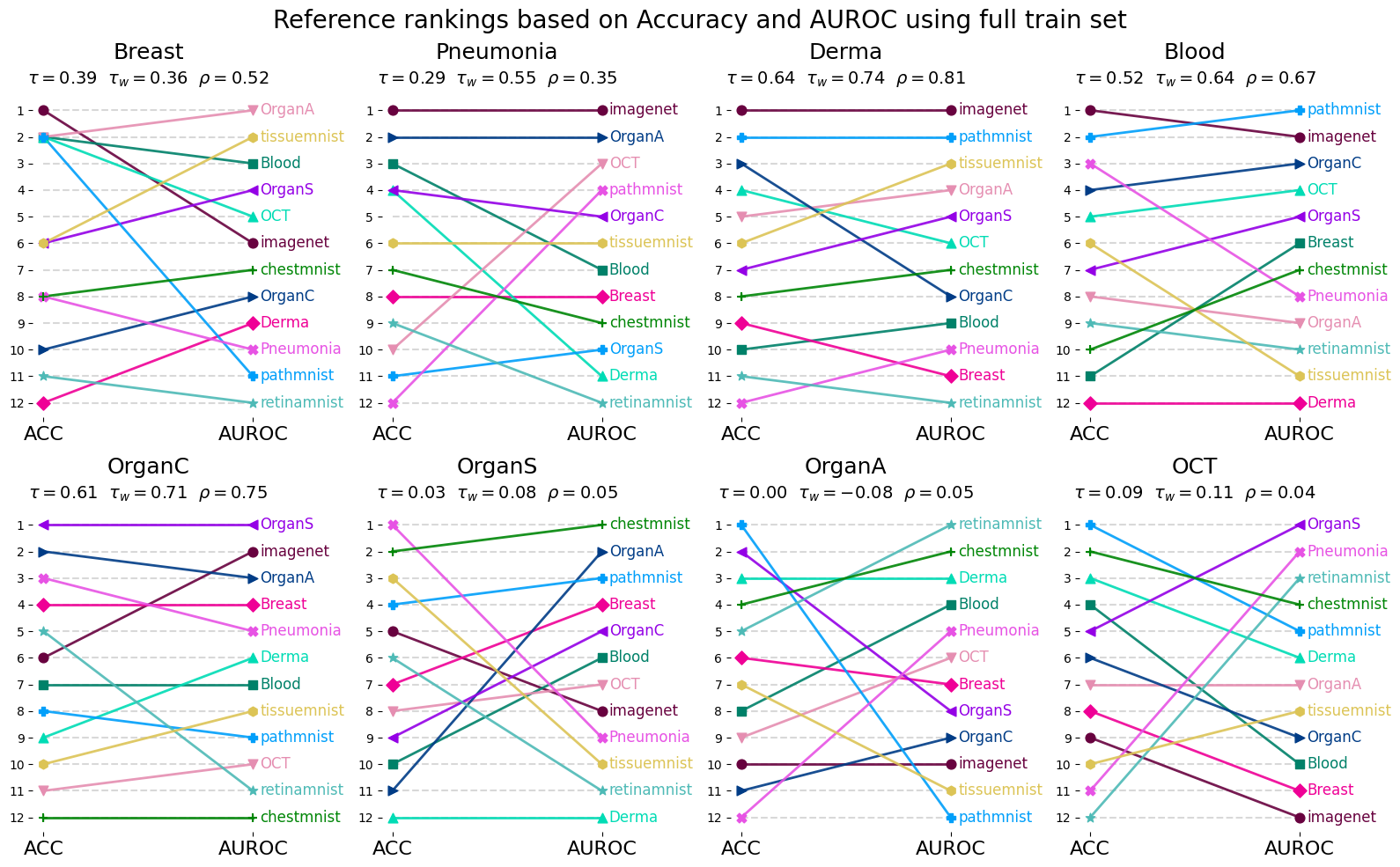}
    \caption{Stability of rankings obtained from two fine-tuning runs, optimized separately for accuracy (ACC) and AUROC on the full train set, considering Kendall's Tau $\tau$, weighted Kendall's Tau $\tau_w$, and Spearman's $\rho$.}
    \label{fig:reference_ranking_alignment_auc_acc_all}
\end{figure}

\begin{figure}[h!]
    \centering
    \includegraphics[width=1\textwidth]{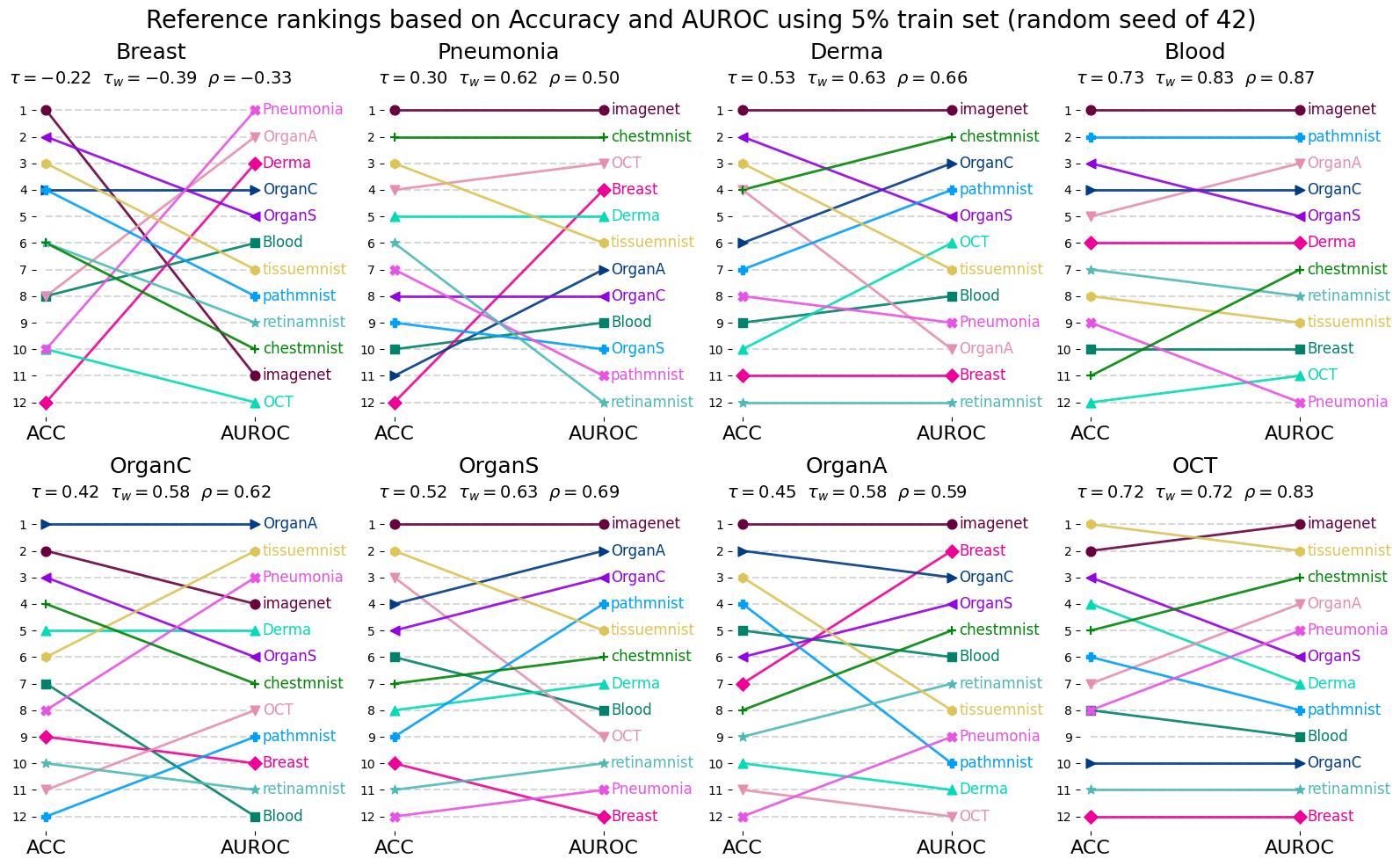}
    \caption{Stability of rankings obtained from two fine-tuning runs, optimized separately for accuracy (ACC) and AUROC on the 5\% of train set (sampled with random seed 42), considering Kendall's Tau $\tau$, weighted Kendall's Tau $\tau_w$, and Spearman's $\rho$.}
    \label{fig:reference_ranking_alignment_auc_acc_5pct_split1_all}
\end{figure}

\begin{figure}[h!]
    \centering
    \includegraphics[width=1\textwidth]{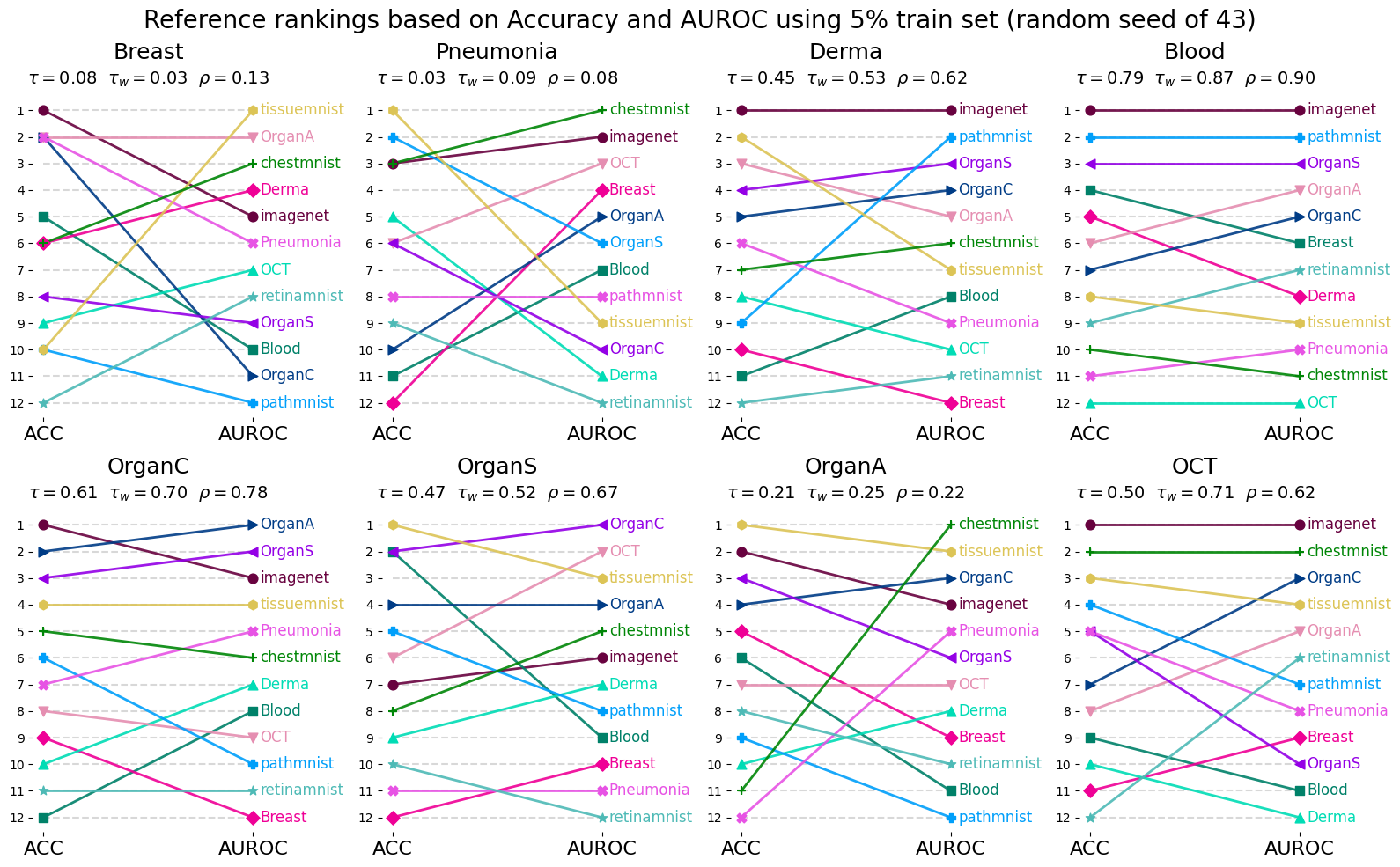}
    \caption{Stability of rankings obtained from two fine-tuning runs, optimized separately for accuracy (ACC) and AUROC on 5\% of train set (sampled with random seed 43), considering Kendall's Tau $\tau$, weighted Kendall's Tau $\tau_w$, and Spearman's $\rho$.}
    \label{fig:reference_ranking_alignment_auc_acc_5pct_split2_all}
\end{figure}

\clearpage
\section{Rank agreement with reference}
\label{appx:stability_ref_from_finetuning}
\renewcommand{\figurename}{G}
\setcounter{figure}{0}

The following figures present the agreement between TE metrics and reference ranking for all target dataset subsets which is described in section \ref{results_reference_ranking}.

\begin{figure}[h!]
    \centering
    \includegraphics[width=1\textwidth]{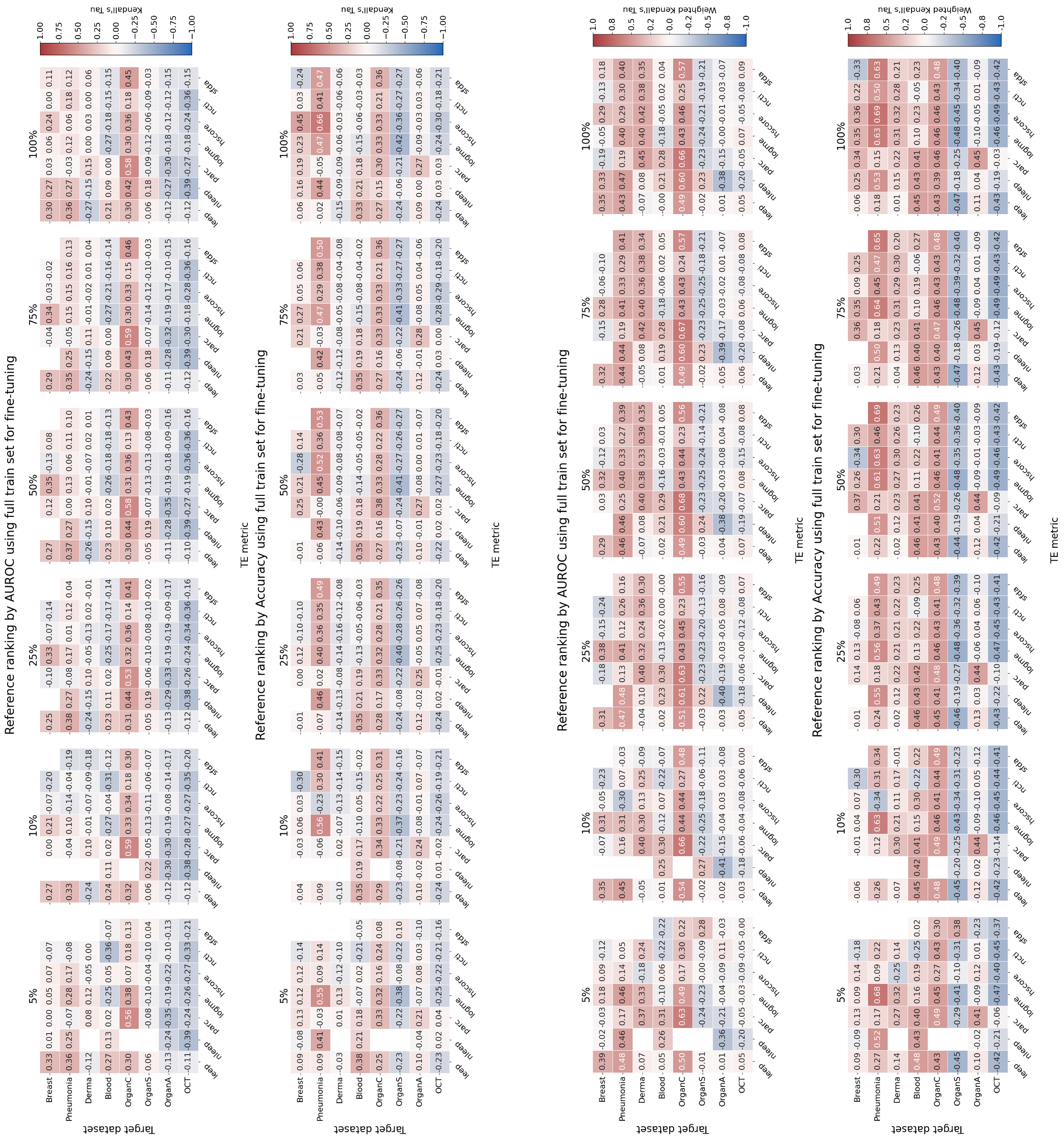}
    \caption{Average pairwise Kendall’s Tau and weighted Kendall's Tau ($stability_{ref}$ ) between TE metrics and reference ranking (averaged over 5 random seeds for subsets < 100\%), optimized separately for accuracy and AUROC.}
    \label{fig:stability_ref_finetuning_all}
\end{figure}

\begin{figure}[h!]
    \centering
    \includegraphics[width=0.5\textwidth]{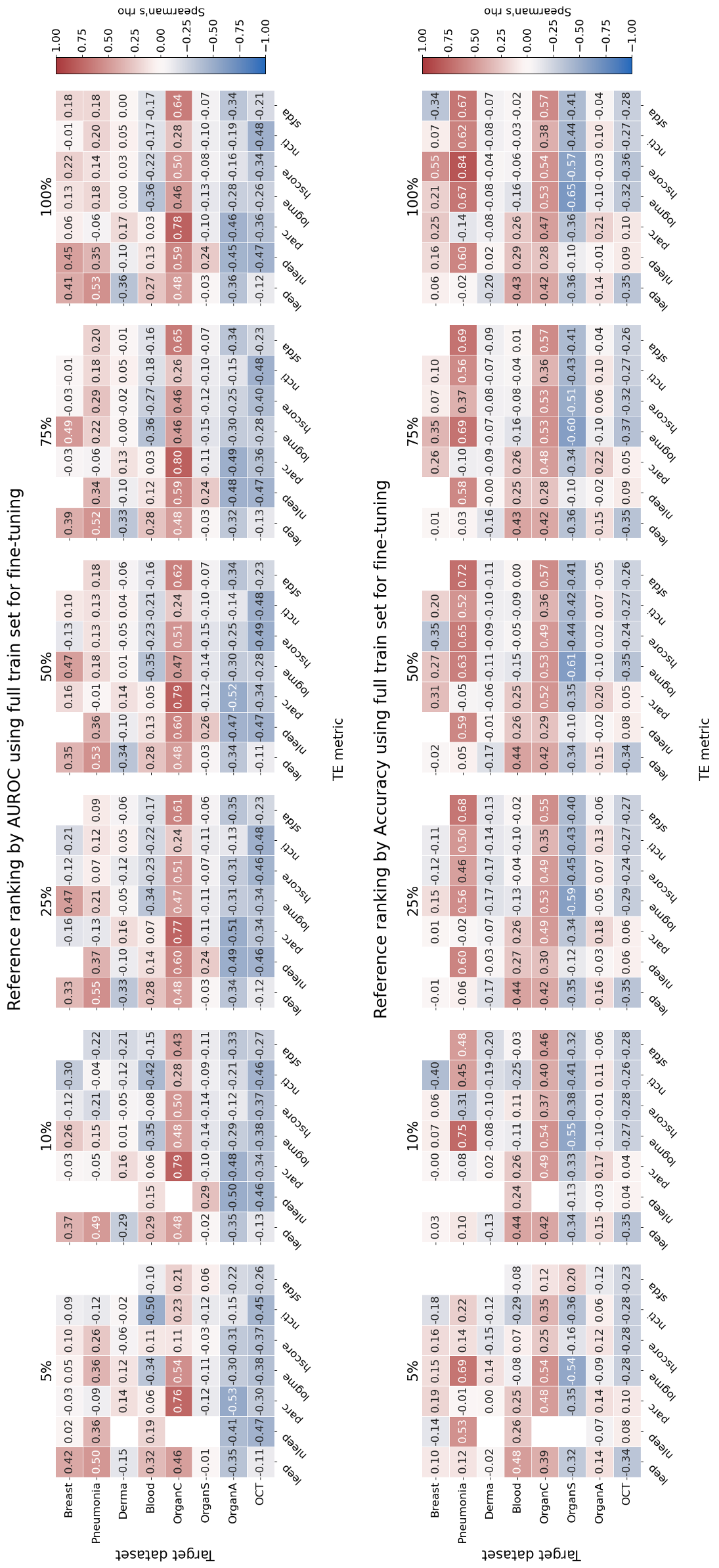}
    \caption{Average pairwise Spearman's rho ($stability_{ref}$ ) between TE metrics and reference ranking (averaged over 5 random seeds for subsets < 100\%), optimized separately for accuracy and AUROC.}
    \label{fig:stability_ref_finetuning_second}
\end{figure}

\end{document}